\documentclass{article}

\usepackage{arxiv}
\usepackage[utf8]{inputenc}
\usepackage[T1]{fontenc}
\usepackage{hyperref}
\usepackage{url}
\usepackage{booktabs}
\usepackage{amsfonts}
\usepackage{amsmath}
\usepackage{nicefrac}
\usepackage{microtype}
\usepackage{cleveref}
\usepackage{graphicx}
\usepackage[numbers]{natbib}
\usepackage{doi}
\usepackage{listings}
\usepackage{xcolor}

\usepackage{algorithmic}
\usepackage{array}
\usepackage[caption=false,font=normalsize,labelfont=sf,textfont=sf]{subfig}
\usepackage{textcomp}
\usepackage{stfloats}
\usepackage{verbatim}
\usepackage{balance}
\usepackage{makecell} 	
\usepackage{multirow}
\usepackage{tabularx}

\title{SupraSNN: Exploiting Synapse-Level Parallelism in Spiking Neural Network Accelerators through Co-Optimized Mapping and Scheduling}

\date{}

\author{ Seyed~Sadra Ghavami~\textsuperscript{\href{https://orcid.org/0009-0008-2719-1384}{\includegraphics[width=8pt]{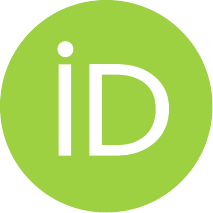}}}\\
	High-Performance Embedded Architecture Laboratory (HiPEAL)\\
	School of Electrical and Computer Engineering\\
	College of Engineering, University of Tehran\\
	Tehran, Iran \\
	\texttt{sadra.ghavami@ut.ac.ir} \\
	\And
	Mohammad~Hossein Nikkhah~\textsuperscript{\href{https://orcid.org/0009-0004-0351-4039}{\includegraphics[width=8pt]{orcid.pdf}}}\\
	High-Performance Embedded Architecture Laboratory (HiPEAL)\\
	School of Electrical and Computer Engineering\\
	College of Engineering, University of Tehran\\
	Tehran, Iran \\
	\texttt{mohammadhossein.nikkhahghomi@epfl.ch} \\
	\And
	Mohammad~Rasoul Roshanshah~\textsuperscript{\href{https://orcid.org/0000-0001-5650-7925}{\includegraphics[width=8pt]{orcid.pdf}}}\\
	High-Performance Embedded Architecture Laboratory (HiPEAL)\\
	School of Electrical and Computer Engineering\\
	College of Engineering, University of Tehran\\
	Tehran, Iran \\
	\texttt{mrroshanshah@ut.ac.ir} \\
	\And
	Saeed Safari~\textsuperscript{\href{https://orcid.org/0000-0001-6940-591X}{\includegraphics[width=8pt]{orcid.pdf}}}\\
	High-Performance Embedded Architecture Laboratory (HiPEAL)\\
	School of Electrical and Computer Engineering\\
	College of Engineering, University of Tehran\\
	Tehran, Iran \\
	\texttt{saeed@ut.ac.ir} \\
}

\renewcommand{\shorttitle}{SupraSNN}

\hypersetup{
	pdftitle={SupraSNN}
}

\begin{document}
	\maketitle
	
	\begin{abstract}
		Spiking Neural Networks (SNNs) offer a brain-inspired path toward highly efficient computation, but their practical deployment is constrained by the challenge of managing and executing their massive parallelism on physical hardware. This problem mirrors the historical challenge in processor design of moving beyond serial execution, a barrier broken by superscalar architectures that dispatch multiple instructions to parallel functional units. Drawing inspiration from this paradigm, we introduce a hardware-software co-design framework that treats synaptic events as parallelizable micro-operations.
		We present SupraSNN, a superscalar-inspired architecture that achieves high synapse-level parallelism by physically decoupling synaptic and neuronal computations. Within this architecture, a Multi-Cast Tree routes spike data to multiple parallel Synapse Processing Units serve as the computational pipelines, while a Merge Tree consolidates distributed results for processing by a unified Neuron Unit--deliberately centralizing complex neuron state dynamics to mitigate hardware overhead and resource duplication.
		The efficacy of this architecture is enabled by a sophisticated partitioning and scheduling framework that first maps the SNN onto hardware respecting memory constraints, then heuristic scheduling determines the synaptic execution order, maximizing throughput and resource utilization.
		Implementing a feedforward SNN trained on MNIST (93.44\% accuracy), SupraSNN achieves 149 $\mu s$ inference latency and 0.025 mJ per image (0.276 nJ per synapse) on the Xilinx Zynq XC7Z020 FPGA--delivering 47.6\% lower latency and 5.6$\times$ better energy efficiency than prior FPGA-based SNN accelerators. Beyond vision tasks, a recurrent SNN on the Spiking Heidelberg Dataset (71.82\% accuracy) achieves 1.41 ms latency and 0.77 mJ per sample on XC7Z030.
	\end{abstract}
	

	\section{Introduction}
	Artificial neural networks have revolutionized computing paradigms for machine learning workloads, yet conventional deep learning architectures face fundamental challenges in efficiently processing temporal, event-based data streams \cite{A. Kuggle et al.}. Spiking Neural Networks (SNNs) have emerged as a promising model for temporal processing and event-based sensing because they natively capture rich spatiotemporal dynamics \cite{snnTorch}. Additionally, by intrinsically exploiting the temporal domain and operating in a highly sparse, event-driven manner, SNNs offer a biologically plausible pathway to extremely energy-efficient computation \cite{W. Mass}.
	
	The actualization of these theoretical benefits depends heavily on the underlying hardware implementation \cite{Are SNNs Truly Energy-efficient}. While SNNs can be simulated on general-purpose processors, specialized hardware is required to truly capture the massive parallelism, microsecond-level latency, and drastic power reductions promised by these networks \cite{C. D. Schuman et al.}.
	
	Consequently, the pursuit of specialized neuromorphic accelerators has led to a broad design space encompassing analog \cite{analog}, \cite{Neuromorphic Electronic Circuits}, mixed-signal \cite{DYNAP}, \cite{Neurogrid}, and fully digital architectures \cite{Loihi}, \cite{TrueNorth}. Digital hardware implementations, in particular, provide a robust solution by offering strict mathematical determinism \cite{C. Li et al.}, high scalability across modern technology nodes \cite{Qiao et al.}, \cite{TrueNorth}, and resilience against the noise and process variations typically found in analog neuromorphic systems \cite{A. Gebregiorgis et al.}. Moreover, digital systems allow for the precise control and reconfigurability necessary to implement the complex, time-dependent dynamics of various rapidly evolving spiking neuron models \cite{Loihi}.
	
	However, existing digital hardware implementations require further improvement to maximize their computational throughput, as handling modern, highly sparse, and irregularly connected SNNs demands deep, fine-grained parallelism \cite{Stitch-X}. Traditional SNN accelerators tightly couple synaptic accumulations with neuronal membrane potential updates \cite{SpiNNaker}, \cite{DYNAP}. This structural rigidity fundamentally limits parallel execution because modern SNN topologies exhibit a severe computational asymmetry: synapses demand a massive number of simple accumulations, whereas neurons require a significantly lower frequency of structurally complex, state-dependent updates \cite{Synapse-Centric Mapper}. To decouple these processes and unlock higher throughput, we draw inspiration from the evolution of high-performance single-core CPUs—specifically, the superscalar microarchitecture \cite{Computer Architecture}, \cite{superscalar}. These architectures achieve high throughput by issuing multiple independent instructions to parallel execution units (ALUs) and subsequently merging the results back into a centralized, deterministic architectural state (the Register File). Guided by this paradigm, we propose a novel architectural approach for digital neuromorphic engines, which we conceptualize as \textit{synapse-level parallelism}. By treating individual synaptic operations as independent tasks that can be dynamically dispatched to parallel units, this approach fundamentally redefines how SNN workloads are executed.
	
	This superscalar-inspired shift toward \emph{synapse-level parallelism} enables three key advantages for modern SNN workloads:
	\begin{enumerate}
	\item \textbf{Controlled Parallelism:} By decoupling synaptic operations from the update cycle of the entire neuron, the engine can achieve a much higher, strictly controlled level of parallelism. Multiple synaptic events can be processed simultaneously across different units, completely independent of the slower membrane potential updates \cite{Spiking Neuron Model}.
	\item \textbf{Irregular Dataflow Flexibility:} It provides the vital flexibility required to handle the irregular connectivity of advanced SNN architectures. This is critical for mapping complex topologies, including fully recurrent and randomly connected layers, which do not follow the rigid dataflow patterns of traditional convolutional networks \cite{SaARSP}.
	\item \textbf{Unstructured Sparsity Support:} It allows the hardware to natively support unstructured weight sparsity with fully general connectivity. Instead of being bound by dense matrix structures, the engine skips zero-valued synapses at a fine-grained level, translating algorithmic sparsity into direct latency and energy savings \cite{EIE}, \cite{Stitch-X}.
	\end{enumerate} 
	
	However, implementing such fine-grained parallelism within an SNN accelerator introduces several non-trivial hardware challenges:
	\begin{itemize}
	\item \textbf{Selective spike distribution (multicasting):} Global broadcasting of spikes to every processing unit simplifies routing logic but results in massive interconnect energy waste. An efficient mechanism is needed to distribute spikes only to the specific parallel units that require them.
	\item \textbf{Parallelism--memory trade-off:} Partitioning synapses widely across parallel compute units exposes abundant parallelism but creates fragmented partial results that must be buffered and merged. Conversely, clustering neuron states to minimize storage leads to severe compute load imbalances under irregular connectivity.
	\item \textbf{Deterministic commit (accumulation):} When multiple synaptic updates targeting the same post-synaptic neuron are produced in parallel, the system must ensure a deterministic commit to the neuron's state. Relying on queues or atomic memory updates can bottleneck high fan-in neurons and introduce non-deterministic mathematical behavior that depends entirely on low-level hardware arbitration.
	\end{itemize}  
	
	To overcome these challenges, we present \textbf{SupraSNN}, an SNN processing engine designed to realize \emph{synapse-level parallelism} through a superscalar-inspired architecture. SupraSNN effectively treats incoming spikes as "instructions" that are dispatched to a parallel array of Synapse Processing Units (SPUs). These SPUs operate in parallel to calculate synaptic contributions, which are then synchronously and deterministically committed to a centralized Neuron Unit utilizing a custom buffer-less merge fabric. This decoupled approach allows our single-core engine to exploit massive parallelism while maintaining the strict mathematical determinism and flexibility required for highly irregular SNN topologies.
	
	The primary contributions of this work are summarized as follows:
	\begin{itemize}
		\item \textbf{Superscalar-Inspired Single-Core Engine} that physically 
		decouples synapse and neuron operations, introducing fine-grained, 
		\emph{synapse-level parallelism} within a single processing core.
		
		\item \textbf{Lightweight Multi-Cast (MC) Tree} that employs an 
		$\mathcal{O}(N)$ scaling bitstream for spike multicasting, strictly avoiding the overhead of global broadcasting 
		and large routing tables.
		
		\item \textbf{Bufferless Merge (ME) Tree} that deterministically and 
		synchronously commits all parallel synaptic partial sums to the 
		centralized neuron state, eliminating queue and atomic lock bottlenecks.
		
		\item \textbf{Hardware-Software Co-Design Framework} that resolves the 
		parallelism--memory trade-off by intelligently partitioning irregular 
		SNN workloads across SPUs.
		
		\item \textbf{Evaluation on Sparse and Temporal Datasets} via FPGA 
		prototyping, demonstrating that the architecture dynamically scales 
		latency and power consumption in direct proportion to unstructured 
		network sparsity, with competitive performance on both the MNIST 
		\cite{MNIST} and Spiking Heidelberg Dataset (SHD) \cite{SHD} benchmarks.
	\end{itemize}
	
	\section{Background}
	\noindent SNNs--often regarded as the third generation of neural networks--extend traditional artificial models by incorporating stateful neurons and adaptive synaptic connections, thereby more closely resembling biological computation \cite{W. Mass}. In the brain, a neuron consists of several main components: the soma (cell body), which maintains the neuron’s membrane potential; the dendrites, which receive spikes (electrical impulses) from other neurons; and the axon, which transmits spikes to neighboring neurons via axon terminals \cite{snnTorch}. When an excitatory synapse receives a spike, it causes a rapid increase in the postsynaptic membrane potential \cite{Spiking Neuron Model}. Once the membrane potential exceeds a defined threshold, the neuron emits a spike and the potential is reset to a reset value \cite{Towards Spike-Based Machine Intelligence}. The generated spike then propagates along the axon and reaches the synaptic terminals, where it is transmitted to the dendrites of neighboring neurons \cite{Deep Learning With Spiking Neurons}. The magnitude and sign of the resulting postsynaptic response depend on the ionic concentration and synaptic strength at the corresponding connection \cite{Spiking Neuron Model}.
	
	To mathematically model this behavior, the Hodgkin–Huxley (HH) \cite{HH} model was introduced as a detailed biophysical representation of neural dynamics, although its computational complexity limits its practical use in large-scale simulations and hardware implementations. The Izhikevich model \cite{Izhikevich} later provided a simplified biologically plausible alternative, yet the Leaky Integrate-and-Fire (LIF) model ultimately emerged as the most practical compromise between biological realism and computational simplicity \cite{Tavanaei et al}. Its simple dynamics, based on a linear differential equation with a threshold-and-reset mechanism, enable low-latency, energy-efficient operation while maintaining sufficient fidelity for large-scale network modeling.
	
	In the LIF neuron, the membrane---representing the internal state of the neuron---is modeled as a capacitor, while the leak mechanism is represented by a resistive path \cite{Spiking Neuron Model}. An external input current is injected through another resistive branch \cite{Spiking Neuron Model}. As a result, the membrane voltage follows the dynamics of a standard
	resistor--capacitor (RC) circuit, expressed in continuous time by \cite{snnTorch}
	\begin{equation}
		\tau_m \frac{dV_m(t)}{dt}
		= -V_m(t) + R_m I(t),
		\label{eq:lif_continuous}
	\end{equation}
	where $V_m(t)$ is the membrane potential, $\tau_m$ is the membrane time constant, $R_m$ is the membrane resistance, and $I(t)$ is
	the synaptic input current.
	
	Digital implementations require a discretized version of the LIF dynamics, evaluated at distinct timesteps. Using a forward Euler method with timestep $\Delta t$, the discrete-time membrane update can be written approximately as \cite{snnTorch}
	\begin{equation}
		V_m^{\mathrm{updated}}[t]
		= (1-\alpha)\, V_m[t] + I[t],
		\label{eq:lif_discrete}
	\end{equation}
	where $\alpha = 1 - e^{\frac{-\Delta t}{\tau_m}}$ is the leak factor.
	
	The synaptic input current at each timestep is computed as a weighted sum of
	presynaptic spikes, \cite{snnTorch}
	\begin{equation}
		I[t] = \sum_{i=0}^{k} W_i\, S_i[t],
		\label{eq:current_discrete}
	\end{equation}
	where $W_i$ is the synaptic weight and $S_i[t] \in \{0,1\}$ is the spike from the
	$i$-th presynaptic neuron at time $t$.
	
	The neuron emits an output spike whenever its updated membrane potential exceeds the
	firing threshold $V_{\mathrm{th}}$: \cite{snnTorch}
	\begin{equation}
		S_{\mathrm{out}}[t] =
		\begin{cases}
			1, & \text{if } V_m^{\mathrm{updated}}[t] \ge V_{\mathrm{th}},\\[4pt]
			0, & \text{otherwise}.
		\end{cases}
		\label{eq:spike_discrete}
	\end{equation}
	Finally, the membrane potential for the next timestep is either reset to the reset
	potential (if a spike is generated) or set to the updated value (if no spike occurs): \cite{snnTorch}
	\begin{equation}
		V_m[t+1] =
		\begin{cases}
			V_{\mathrm{reset}}, & \text{if } V_m^{\mathrm{updated}}[t] \ge V_{\mathrm{th}},\\[4pt]
			V_m^{\mathrm{updated}}[t], & \text{otherwise}.
		\end{cases}
		\label{eq:vm_reset}
	\end{equation}
	
\begin{figure}
		\centering
		\includegraphics[width=\linewidth]{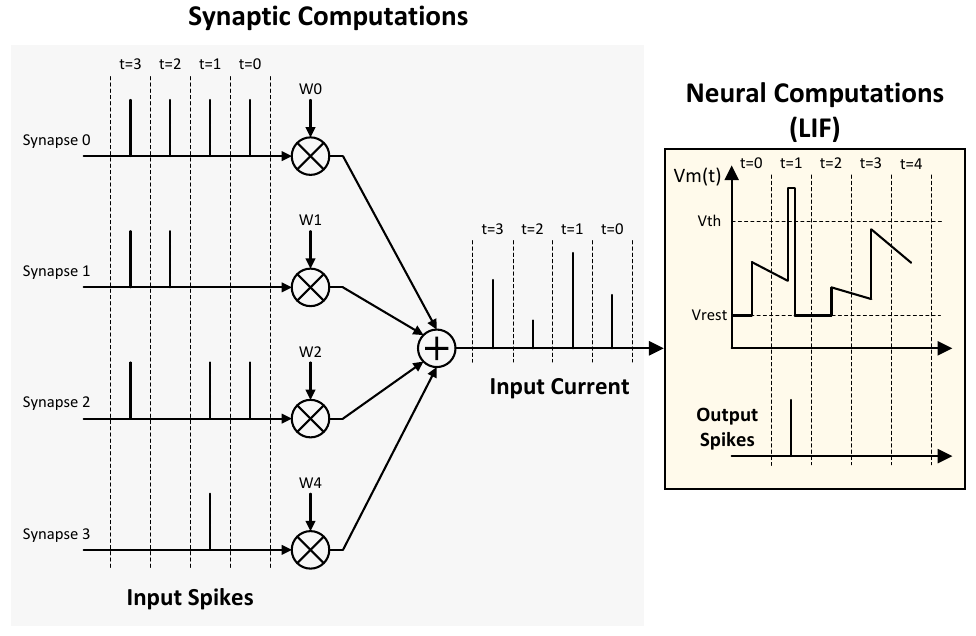}%
		\caption{Synaptic and neuronal computations in a discrete-time LIF neuron. 
			Presynaptic input spikes are weighted and accumulated to form the input current,
			which is then integrated by the LIF membrane dynamics to produce output spikes.}
		\label{fig:lif}
	\end{figure}

	In summary, each timestep of the LIF neuron involves:
	(i) \emph{synaptic computation}, where the input current $I[t]$ is obtained by
	accumulating weighted presynaptic spikes according to~\eqref{eq:current_discrete};
	and (ii) \emph{neuronal computation}, where the membrane potential is updated via
	leaky integration, compared against the threshold, and either reset or carried forward
	according to~\eqref{eq:lif_discrete}--\eqref{eq:vm_reset}.
	Figure~\ref{fig:lif} illustrates this complete process.
	
\begin{figure}
		\centering
		\includegraphics[width=\linewidth]{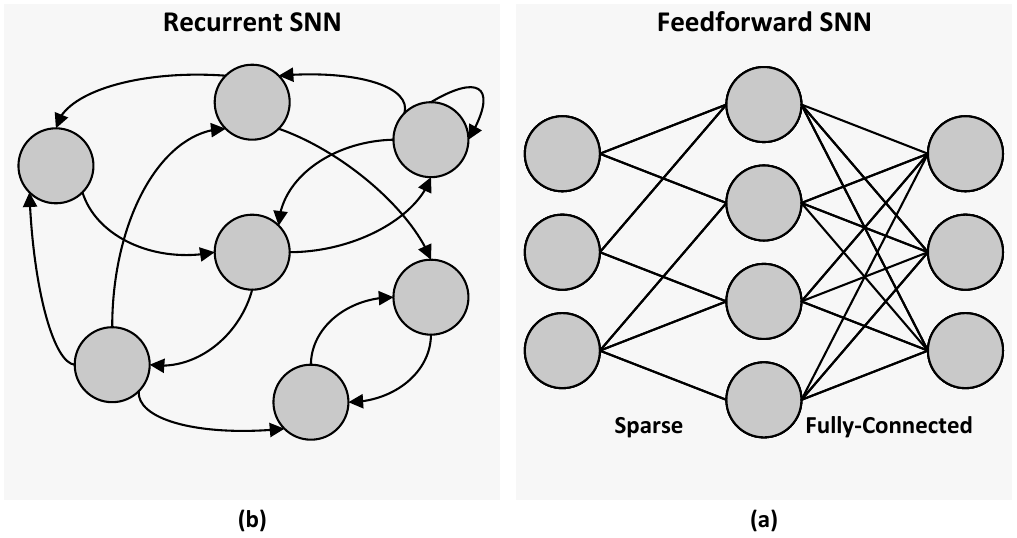}%
		\caption{Common connectivity schemes in SNNs.
			(a) Feedforward SNNs with either sparse or fully-connected layers.
			(b) Recurrent SNN, representative of reservoir or liquid-state topologies.}
		\label{fig:topologies}
	\end{figure}

	The most common structure is the Spiking Feedforward Neural Network (SFNN), in which neurons in each layer receive spikes only from the preceding layer \cite{Tavanaei et al}, with connections being either fully connected or sparse, as illustrated in Figure ~\ref{fig:topologies}a. Such architectures are typically used in pattern-recognition tasks, including classification of static sensory inputs \cite{Deng et al.}.
	
	Beyond feedforward architectures, the Spiking Recurrent Neural Network (SRNN) is another important structure, in which neurons form a sparse, irregular recurrent graph with randomized connectivity \cite{snnTorch}, as shown in Figure ~\ref{fig:topologies}b. Such networks are particularly well suited for temporal processing, including speech recognition, gesture sequence classification, and general time-series analysis \cite{Spiking Neural Networks: A Survey}.
	
	\section{Related Work}
	\noindent A wide range of neuromorphic processors and accelerators have been proposed to enable low-power inference and learning in SNNs. One representative design is ODIN \cite{ODIN}, a fully digital neuromorphic processor capable of simulating 256 neurons with 64k synapses and supporting on-chip online learning through the Spike-Driven Synaptic Plasticity (SDSP) rule. ODIN employs a time-multiplexed crossbar organization, in which synapse and neuron update logic are shared over time to emulate a large synaptic array while significantly reducing area and power overhead. Its architectural organization is structured into three conceptual stages—analogous to dendrites, soma, and axon—corresponding respectively to excitatory/inhibitory weight accumulation, subthreshold neuron dynamics, and event generation. ODIN also extends the Address-Event Representation (AER) \cite{AER} protocol to support multiple event types, including neuron spikes and time-reference events, enabling fully event-driven computation. A key component is its internal event scheduler, which manages single-spike and burst events using a combination of a 32-stage FIFO and multiple rotating 4-stage FIFOs. This event-centric architecture supports both inference and online learning with high energy efficiency. However, the time-multiplexed crossbar requires updating the full set of postsynaptic entries for each presynaptic spike, which may introduce scalability challenges when network connectivity is sparse, due to many SRAM accesses that do not contribute to computation.
	
	A contrasting approach is demonstrated by Spiker+ \cite{Spiker+}, a configurable hardware–software framework that automatically generates FPGA-based SNN accelerators from high-level network descriptions. In Spiker+, each neuron in a layer is mapped to a lightweight physical neuron core implementing variants of the LIF model, enabling high degrees of parallelism for fully connected feedforward and recurrent networks. The control flow is intentionally simple and hierarchical: neuron-level, layer-level, and network-level control units coordinate computation through a two-signal start/ready handshake. Spikes are propagated through parallel inter-layer connections, allowing clock-driven real-time inference on small workloads. The architecture is designed to minimize logic complexity and power consumption; however, its fully parallel neuron-per-neuron mapping requires accessing all synaptic weights in parallel at each timestep, making on-chip memory capacity the primary scalability constraint. This resource requirement limits the practical size of networks that can be deployed on FPGA devices.
	
	ODIN and Spiker+ represent two opposing extremes of a fundamental design trade-off: minimizing hardware resources at the cost of throughput, or maximizing parallelism at the cost of scalability. Neither addresses the intervening design space of balancing fine-grained parallel execution with efficient resource utilization under irregular connectivity--precisely the gap that motivates the architecture proposed in this work.
	
	To exploit parallelism in neuromorphic systems, partitioning, mapping, and scheduling are essential steps. SpiNeMap \cite{SpiNeMap} is a prominent work addressing this problem for multi-crossbar neuromorphic hardware. It provides two stages—SpiNeCluster and SpiNePlacer—to partition and map large SNNs onto crossbar arrays connected via a shared interconnect. SpiNeCluster partitions the network into clusters that fit within individual crossbars while minimizing the number of spikes communicated on global synapses—synapses that connect two different clusters—since they generate inter-crossbar traffic, increase spike latency, and consume energy. To achieve this, SpiNeCluster uses a heuristic inspired by the Kernighan–Lin graph-partitioning method. SpiNePlacer then determines where clusters should be placed on the physical crossbars. Placement affects energy and latency because spikes may need to traverse multiple hops in a NoC to reach their target crossbar. SpiNePlacer uses particle-swarm optimization (PSO) combined with Noxim++, an extended cycle-accurate NoC simulator, to evaluate hop counts, routing behavior, congestion, and energy. This two-phase strategy allows SpiNeMap to jointly reduce interconnect energy and spike latency. However, SpiNeMap does not model cluster-level workload balance: some clusters may contain significantly higher firing rates than others, which can create per-crossbar bottlenecks and reduce achievable parallel throughput.
	
	A different perspective is taken in the synapse-centric mapper for SpiNNaker systems \cite{Synapse-Centric Mapper}. SpiNNaker \cite{SpiNNaker} consists of many interconnected chips, each containing 18 ARM cores with limited Tightly Coupled Memory (TCM)—32 KB for instructions and 64 KB for data—and shared SDRAM accessed via DMA. Traditional SpiNNaker mapping assigns post-synaptic neurons to cores, with each core performing both synaptic and neuronal computation for its assigned neurons \cite{PACMAN}. In sparse networks, however, this neuron-centric partitioning becomes inefficient because DMA fetches fan-out–organized rows from SDRAM, and short sparse rows force repeated DRAM accesses, increasing latency and reducing throughput. The synapse-centric mapper restructures the computation by splitting synaptic and neuronal processing across separate cores. Some cores become synapse cores, each responsible for a subset of synaptic rows, while others act as neuron cores that perform membrane updates. Synaptic rows are distributed among synapse cores in a workload-aware way that accounts for firing activity, ensuring more balanced load and better utilization. This design increases the efficiency of DRAM row fetches because every fetched row is fully consumed by the assigned synapse core. However, separating synapse computation introduces a partial-sum merging overhead: neuron cores must retrieve partial contributions from SDRAM and accumulate them, which can become a communication bottleneck and limit achievable parallelism.
	
	Together, these two frameworks indicate a gap. SpiNeMap \cite{SpiNeMap} focuses on reducing interconnect cost (global-synapse traffic and placement-induced latency), whereas the synapse-centric mapper \cite{Synapse-Centric Mapper} focuses on core-level workload balance and efficient memory access. Neither framework simultaneously addresses balanced distribution of synaptic workload across parallel compute units and efficient merging of partial neuronal results without introducing bottlenecks, which could improve the resource utilization significantly. This highlights the need for a mapping strategy capable of distributing synchronous synaptic load evenly while also supporting a scalable mechanism for combining partial results in parallel.
	 
	\section{SupraSNN Hardware Architecture}
	\subsection{Design Overview}
\begin{figure}
		\centering
		\includegraphics[width=\linewidth]{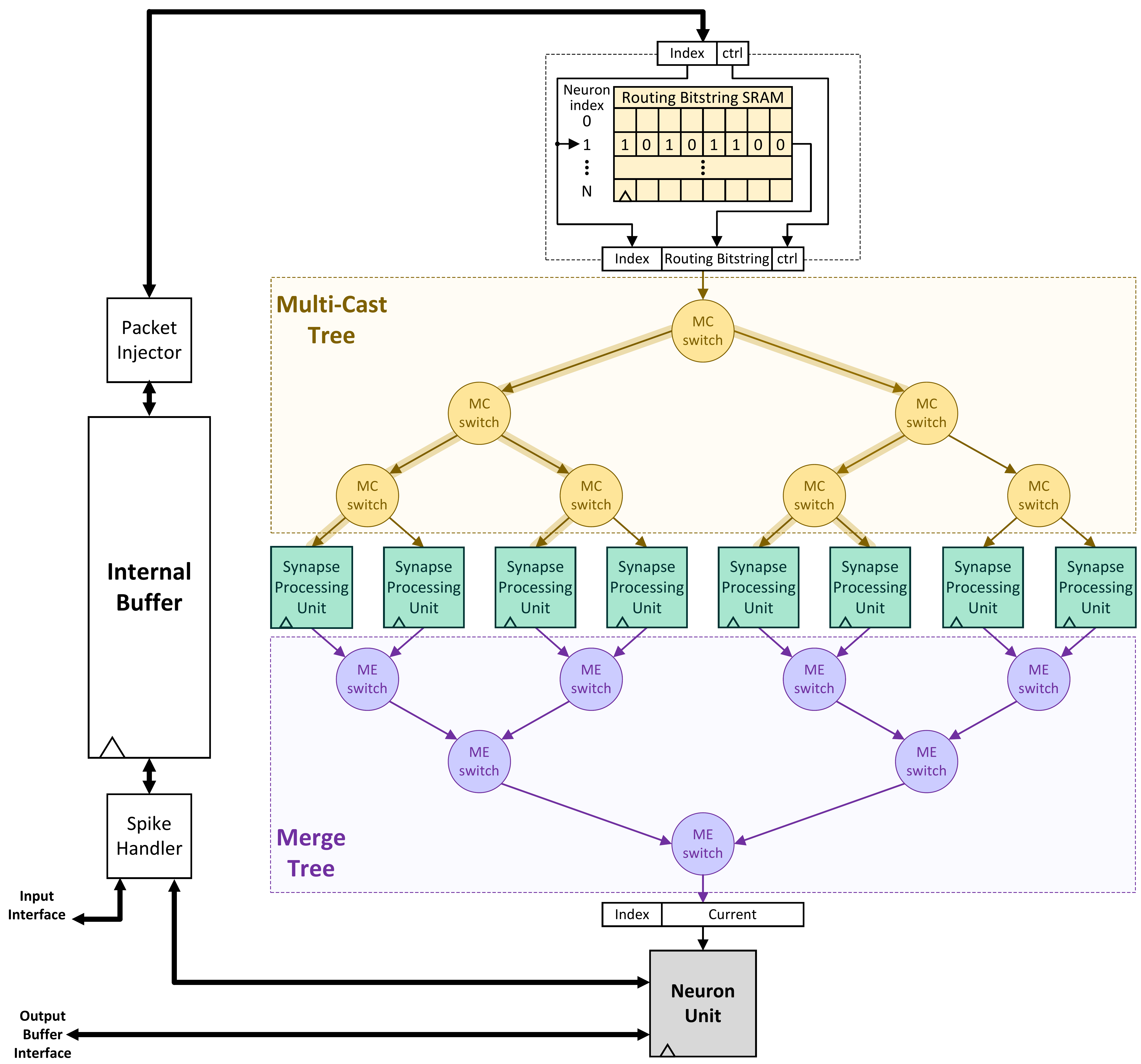}%
		\caption{Overview block diagram of SupraSNN Hardware Architecture.}
		\label{fig:overview}
	\end{figure}

	\noindent Figure~\ref{fig:overview} illustrates the overall architecture of SupraSNN, composed of five principal components. The parallel Synapse Processing Units (SPUs) receive spike events and concurrently execute their assigned synaptic operations. The Routing Unit stores a per-neuron bitstring that encodes which SPUs demand that neuron's spike information. The MC Tree forms a binary dispatch fabric that propagates incoming spikes exclusively to the relevant SPUs, guided by the routing bitstring. The ME Tree serves as a bufferless accumulation fabric that sums the partial results produced by all SPUs in a synchronized manner. Finally, the centralized Neuron Unit receives the fully accumulated synaptic results, updates each neuron's membrane state according to the LIF dynamics, and generates output spikes for the next timestep.
	Drawing inspiration from superscalar processors, these components collectively provide high levels of \emph{synapse-level parallelism} while maintaining deterministic neuron-state updates. In this architecture, synaptic computations are distributed across multiple parallel units, whereas neuronal state is maintained in a single shared unit, analogous to centralized architectural state in superscalar microarchitectures.
	
	\subsection{Data/Execution Flow Explanation}
	\label{subsec:dataflow}
\begin{figure}
		\centering
		\includegraphics[width=\linewidth]{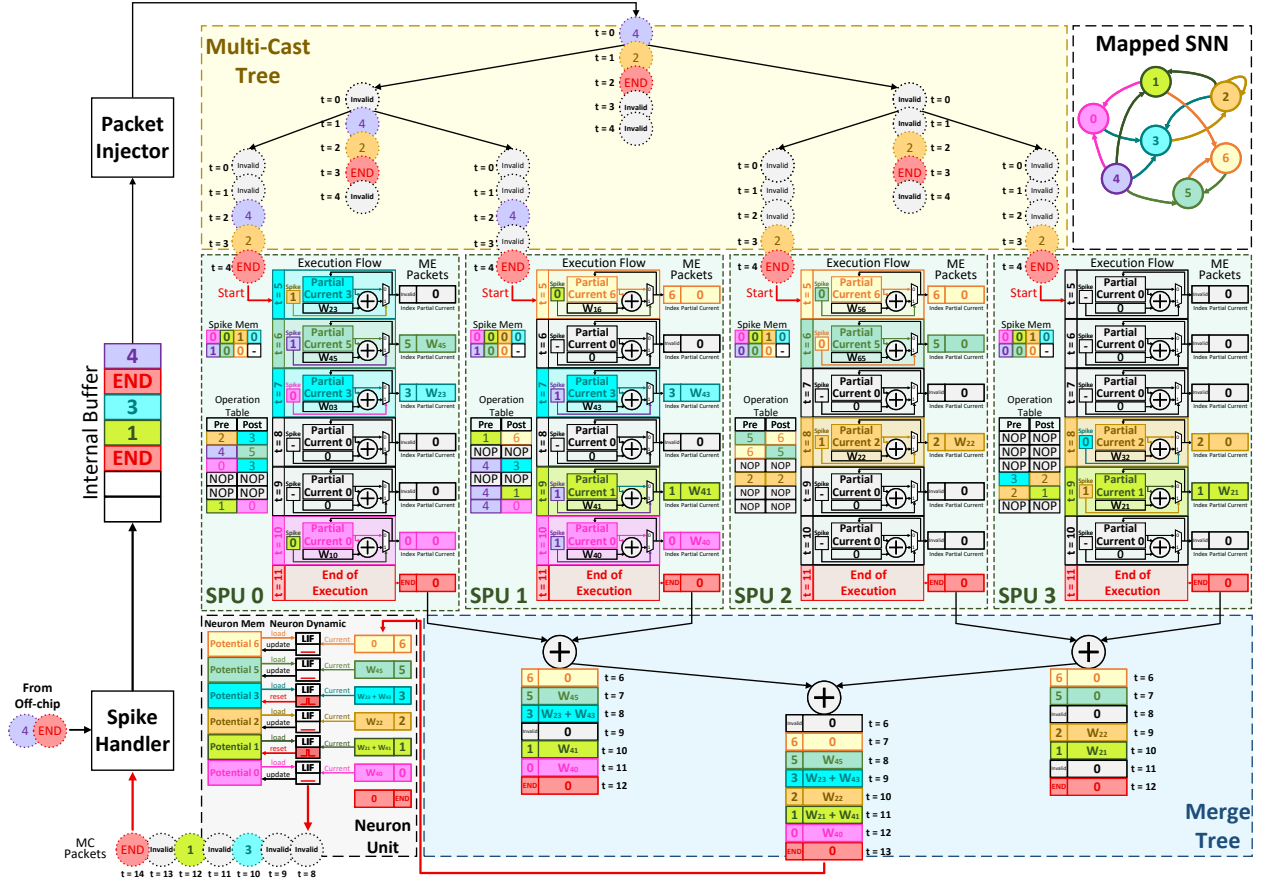}%
		\caption{Abstract representation of the execution dataflow of a sample SNN on SupraSNN for one timestep.}
		\label{fig:timeline}
	\end{figure}

	\noindent Figure~\ref{fig:timeline} illustrates the execution dataflow of a simple SNN mapped onto the SupraSNN architecture during a single simulation timestep. The corresponding network topology, shown in the top-right corner of the figure, consists of seven neurons (indexed 0 to 6), where neuron 4 serves as an input neuron receiving spikes from off-chip. During the mapping stage, all synaptic connections are partitioned across the SPUs and stored in local \emph{Operation Tables}, enabling parallel synaptic operations across distinct hardware units. Each entry encodes essential synaptic parameters, including pre- and post-synaptic neuron indices. For instance, the first entry of SPU 0 stores the synapse from neuron 2 to neuron 3 (synapse 2$\rightarrow$3).
	
	Execution begins with the distribution of spike packets generated during the previous timestep. These AER-inspired \cite{AER} MC packets contain the source neuron index and are initially held in the Internal Buffer. Once all SPUs signal readiness, the Packet Injector sequentially transmits these packets into the MC Tree. In this example, neurons 4 and 2 fired in the preceding timestep; their corresponding MC packets are injected into the tree at $t=0$ and $t=1$, respectively. Here, $t$ represents abstract execution step rather than concrete clock cycles, serving only to illustrate the ordering of execution
	within the architecture.
	
	The MC Tree routes these packets exclusively to targeted SPUs. As shown in Figure~\ref{fig:timeline}, the root node routes the packet for neuron 4 solely toward the left subtree, where it is duplicated to reach SPU 0 and SPU 1 at $t=2$. Upon receiving a packet, each SPU registers the event in its local \emph{Spike Memory} by asserting the corresponding neuron bit, holding it until the end of the current timestep.
	
	After distributing all spike events, the Packet Injector broadcasts an \emph{end packet} as a synchronization barrier. Upon receiving this packet, all SPUs concurrently begin executing the synaptic operations ordered within their local \emph{Operation Tables}. If a pre-synaptic neuron's bit is active in the \emph{Spike Memory}, the SPU accumulates the corresponding synaptic weight into a local partial current assigned to the target post-synaptic neuron. Because a neuron’s total fan-in may be spatially partitioned across multiple SPUs, these distributed partial contributions must be aggregated prior to the state update phase. For example, at $t=9$, SPU 1 and SPU 3 concurrently generate partial currents $W_{41}$ and $W_{21}$ for post-neuron 1.
	 
	An SPU forwards its accumulated partial current to the ME Tree immediately after processing its final synapse for a given post-synaptic neuron. For example, synapses 2$\rightarrow$3 and 0$\rightarrow$3 are both mapped to SPU 0. Although synapse 2$\rightarrow$3 is computed at $t=5$, SPU 0 wait until synapse 0$\rightarrow$3 completes at $t=7$, at which point the partial current for post-neuron 3 is injected into the ME Tree. The ME Tree aggregates distributed partial currents without intermediate hardware buffering, a feature sustained by a compile-time scheduler that synchronizes all SPU output injections for a given post-neuron. SPUs lacking synapses to a currently processing post-synaptic neuron perform a No-Operation (NOP) and transmit an \emph{invalid packet} to maintain pipeline synchronization. 
	 
	To see this merging mechanism in action, consider post-neuron 1. At $t=9$, SPU 1 and SPU 3 inject their valid partial currents $W_{41}$ and $W_{21}$ into the ME Tree, while SPU 0 and SPU 2 simultaneously emit \emph{invalid packets}. At intermediate nodes of the tree, valid partial currents propagate forward unaffected when merged with \emph{invalid packets}. These wavefronts converge at the root of the ME Tree at $t=10$, delivering the fully resolved input current $W_{41} + W_{21}$ to the Neuron Unit at $t=11$. Following the completion of all local synaptic operations, each SPU issues an \emph{end packet} into the ME Tree. These signals converge into a single global synchronization packet at the root node, indicating that all synaptic current accumulations for the timestep are finalized.
	
	The unified Neuron Unit receives these completed input currents and sequentially updates the state variables for all neurons. Utilizing a discrete-time LIF model, it computes the new membrane potential from the previously stored state and the incoming current. If the updated potential exceeds the firing threshold, the Neuron Unit generates an MC packet containing the firing neuron's index and writes it into the Internal Buffer; in this trace, neurons 3 and 1 register spikes at $t=10$ and $t=12$, respectively. Once the \emph{end packet} arrives from the ME Tree, the Neuron Unit forwards it to the Internal Buffer. Concurrently, off-chip input spikes are caught by the Spike Handler and placed in the Internal Buffer, capped by an external \emph{end packet}.
	 
	The timestep concludes when the Packet Injector encounters two distinct \emph{end packets} in the Internal Buffer: one from the Neuron Unit signaling the completion of internal state updates, and one from the Spike Handler indicating the end of external inputs. Upon detecting both barriers, the Packet Injector issues a final \emph{end packet} to the MC Tree, advancing the global system state to the subsequent simulation timestep.
	
	\subsection{MC and ME Communication Fabric}
\begin{figure}
		\centering
		\includegraphics[width=\linewidth]{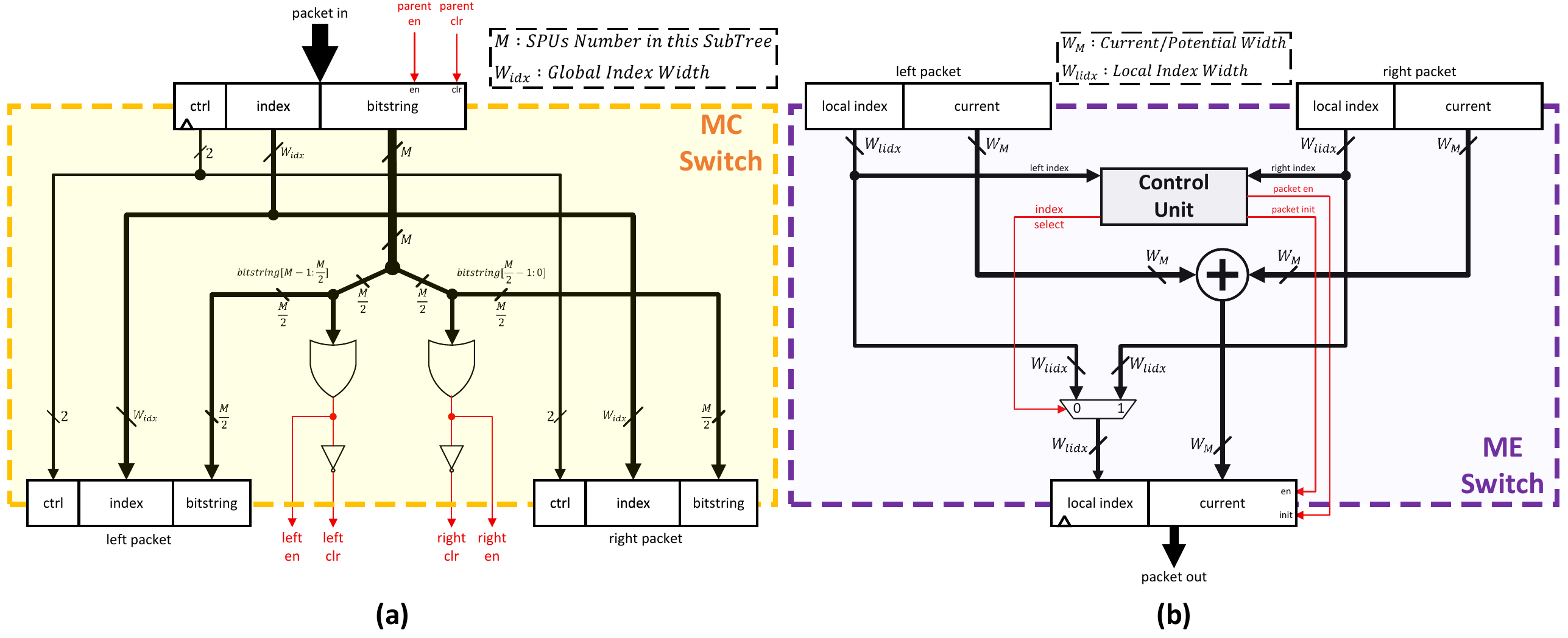}%
		\caption{(a) Micro architecture of a MC switch (b) Micro architecture of a ME switch.}
		\label{fig:Switches}
	\end{figure}

	\noindent To realize a lightweight multicast fabric, SupraSNN appends a routing bitstring to each MC packet, explicitly specifying the target SPUs. A programmable SRAM within the Routing Unit stores one bitstring per neuron, where each bit corresponds to one SPU and indicates whether that SPU holds any synapse originating from that neuron. To utilize this bitstring for routing, as illustrated in Figure~\ref{fig:Switches}a, each MC switch divides it into two equal halves corresponding to the left and right sub-trees and applies a simple OR reduction to each half. The MC packet is forwarded to every direction whose associated half contains at least one set bit. This design avoids the overhead of embedding large routing tables inside switches, which would require $2N(M-1)$ bits of routing memory for $N$ neurons and $M$ SPUs, compared with only $NM$ bits required by the bitstring approach--a clear advantage for any practical design employing more than two SPUs.
	
	Beyond spike delivery, SPUs store synaptic attributes--including pre- and post-synaptic neuron indices and synaptic weights--which must be initialized prior to execution. SupraSNN extends the MC packet format to support high-bandwidth, packet-based initialization through the MC Tree itself, reusing the communication fabric to significantly reduce initialization latency. Only small configuration fields are programmed via conventional bitstreams.
	
	Each MC packet contains a 2-bit control header (\texttt{ctrl}) and a data payload, enabling four distinct packet types summarized in Table~\ref{tab:mc_packet_types}.
	
	\begin{table}
		\centering
		\caption{MC packet types}
		\label{tab:mc_packet_types}
		\begin{tabular}{c l}
			\hline
			\texttt{ctrl} & Payload description \\
			\hline
			00 & Invalid data \\
			01 & Index of neurons that spiked in the last timestep \\
			10 & Index of unit targeted for initialization \\
			11 & Initialization data for the selected unit \\
			\hline
		\end{tabular}
	\end{table}
	
	Packets with \texttt{ctrl} = 00 carry invalid data and allow the MC Tree to operate synchronously without handshaking, simplifying switch logic and reducing control overhead. When \texttt{ctrl} = 01, the payload contains the index of a neuron that generated a spike in the previous timestep. For initialization, packets with \texttt{ctrl} = 10 broadcast the target unit index to all programmable units--including SPUs, the routing bitstream unit, and the Neuron Unit--placing the matching unit into an initialization state. Subsequent packets with \texttt{ctrl} = 11 deliver the corresponding initialization data, which the selected unit consumes while others ignore.
	
	The ME Tree combines partial synaptic results generated by different SPUs through a simple bufferless adder-tree structure. Rather than employing complex buffering switches that wait for all partial results of a neuron to arrive--which incurs substantial area and power overhead--SupraSNN restricts SPUs to a pre-determined injection order so that all partial results for each neuron are injected simultaneously, eliminating buffering entirely. As illustrated in Figure~\ref{fig:Switches}b, each ME switch receives two packets per cycle from its left and right subtrees. If both packets carry the same neuron index, their currents are accumulated and forwarded as a single packet. If either input carries an invalid index, the valid packet is forwarded unchanged. 
	
	To enable seamless communication across the MC and ME fabrics as explained in the Section~\ref{subsec:dataflow}, two reserved indices are defined: the \emph{end index} and the \emph{invalid index}. The \emph{end index} is encoded with all bits set to one and serves as a termination marker in the dataflow. The \emph{invalid index} follows the same encoding except that its least significant bit is cleared to zero.
	
	The paired multicasting-and-merging fabric draws inspiration from the Distribution Tree and Augmented Reduction Tree of MAERI \cite{MAERI}, originally proposed for flexible DNN dataflow mapping, and adapts this paradigm to SNN-specific requirements including spike-based selective multicasting via routing bitstrings and bufferless synchronous partial current merging.
	
	\subsection{Parallel Synapse Processing Units (SPUs)}
\begin{figure}
		\centering
		\includegraphics[width=\linewidth]{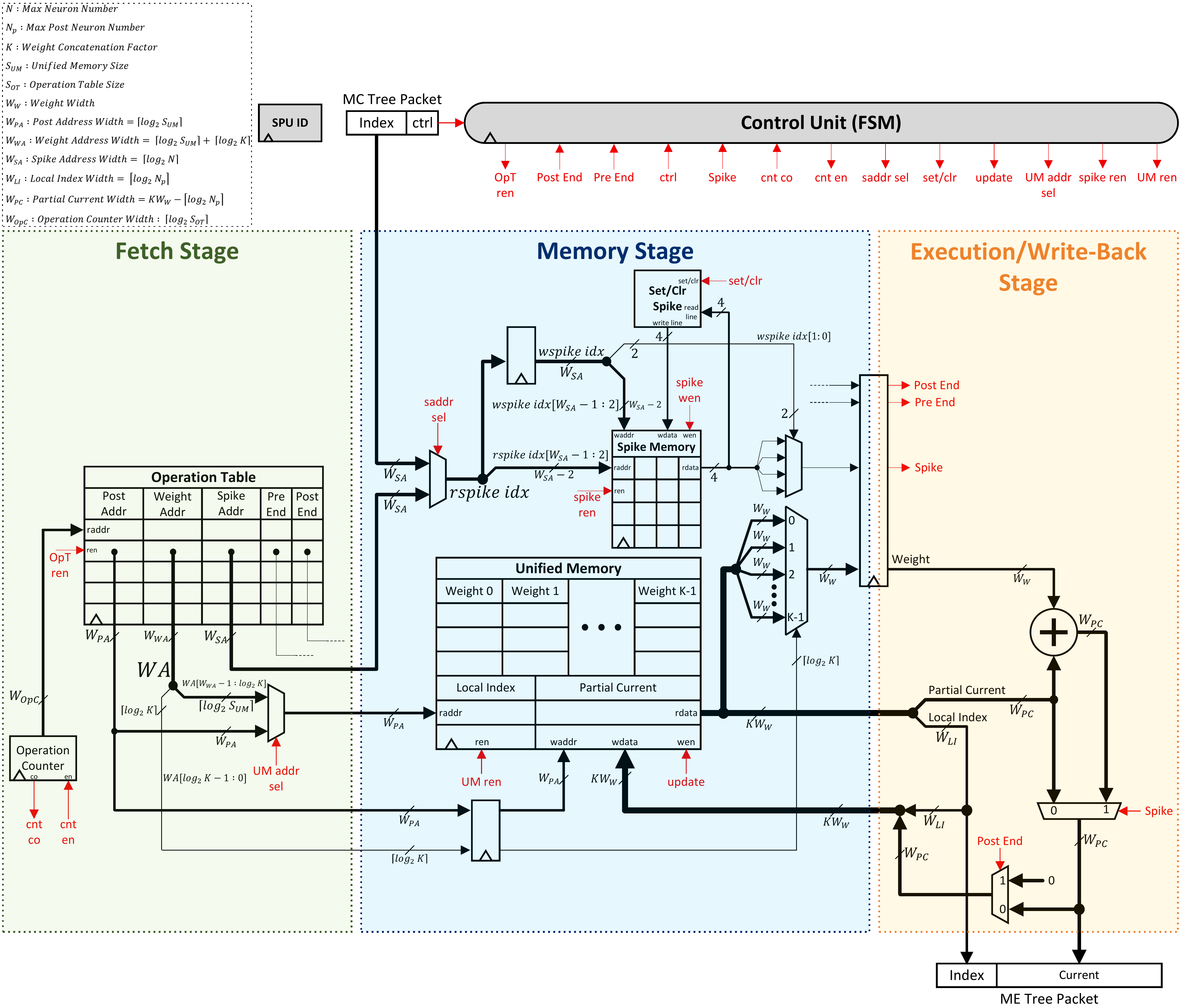}%
		\caption{Micro-architecture of SPU.}
		\label{fig:SPU}
	\end{figure}

	\noindent Figure~\ref{fig:SPU} illustrates the architecture of a SPU. SPUs are responsible for performing synaptic computations in parallel and are designed to provide programmability, energy efficiency, and full flexibility with respect to network connectivity.
	
	\subsubsection{Spike Storage and Event Handling}
	\noindent Synaptic processing in each SPU begins with the \emph{Spike Memory}, a bitmap SRAM that records the spiking activity of the previous timestep as conveyed by MC packets. Each neuron is represented by a single bit indicating whether it fired in the last timestep. The memory is organized into rows of $W_{SM}$ bits (set to 4 in Figure~\ref{fig:SPU}). To simplify addressing, $W_{SM}$ is chosen as a power of two: the $\log_2 W_{SM}$ least significant bits of the neuron index determine the bit position within a row, while the remaining index bits select the corresponding row. When an MC packet arrives, the \emph{Set/Clear Spike} unit sets the corresponding bit in \emph{Spike Memory} to reflect the received spike event. This bitmap organization scales as $\mathcal{O}(N)$ in total memory, avoiding the $\mathcal{O}(N \log N)$ memory burden that buffering neurons' indices would otherwise impose.
	
	Reception of an MC packet carrying the \emph{end index} indicates that all spike events for the current timestep have been delivered. At this point, the SPU can safely assume that \emph{Spike Memory} contains a complete and consistent snapshot of presynaptic activity and can begin synaptic computation (as explained in section~\ref{subsec:dataflow} and shown in Figure~\ref{fig:timeline}).
	
	\subsubsection{Pipeline Organization and Operation Table}
	\noindent Each SPU employs a three-stage pipeline composed of the \emph{Fetch}, \emph{Memory}, and \emph{Execution/Write-Back} stages. During the Fetch stage, an \emph{Operation Counter} iterates through entries in the \emph{Operation Table}, which stores synaptic attributes in an indirect form by storing addresses rather than explicit data values. Each entry in the table specifies the information required to process a single synaptic operation and contains the following five fields:
	\begin{itemize}
	\item \textbf{Post Addr}: the address of the post-synaptic neuron state in \emph{Unified Memory},
	\item \textbf{Weight Addr}: the address of the synaptic weight in \emph{Unified Memory},
	\item \textbf{Spike Addr}: the address of the corresponding pre-synaptic neuron,
	\item \textbf{Pre End}: a flag indicating that this is the final operation associated with the given pre-synaptic neuron for the current timestep, enabling the reset of the corresponding bit in \emph{Spike Memory} for the next timestep,
	\item \textbf{Post End}: a flag indicating that this is the final operation associated with the given post-synaptic neuron for the current timestep.
	\end{itemize}
	
	The proposed \emph{operation-based} implementation provides several architectural advantages:
	\begin{enumerate}
	\item \textbf{Mitigating Weight-Sparsity Overhead:} Zero-weight synapses do not contribute to the postsynaptic neuron state and therefore have no effect on the final simulation results. In the proposed \emph{operation-based} scheme, zero-weight synapses are simply omitted from the \emph{Operation Table}. Consequently, both execution time and memory footprint are reduced.
	\item \textbf{Weight Reusability:} The architecture stores each unique weight only once in the \emph{Unified Memory}. Synapses sharing the same weight reference it through the \textbf{Weight Addr} field in the \emph{Operation Table}, enabling efficient weight sharing across multiple synapses without duplication. Furthermore, for tasks requiring larger weight precision, referencing weights through compact addresses rather than storing explicit values substantially reduces the overall memory footprint.
	\item \textbf{Support for Irregular Connectivity:} By explicitly storing the addresses of both pre- and post-synaptic neurons, the \emph{operation-style} representation naturally supports arbitrary connectivity patterns. This flexibility enables the architecture to efficiently accommodate irregular network topologies commonly found in modern and advanced SNN models.
	\item \textbf{Overlapping Synaptic Computation and Merging:} As illustrated in Figure~\ref{fig:timeline}, the architecture increases throughput by overlapping synaptic computation with current merging in the ME Tree. Each partial current generated by an SPU is forwarded to the ME Tree immediately after it is produced, facilitated by the \textbf{Post End} flag, which marks the final synapse associated with a particular post-synaptic neuron within that SPU. Since the ME Tree operates without internal buffering, the scheduling framework exploits the programmable execution order provided by the \emph{Operation Table} to guarantee that partial currents targeting the same post-neuron are injected simultaneously across all SPUs.
	\end{enumerate}

	\subsubsection{Memory Access and Unified Memory}
	\noindent In the Memory stage, the \textbf{Spike Addr} field is used to read the corresponding bit from the \emph{Spike Memory}, indicating whether the pre-synaptic neuron spiked in the previous timestep. In parallel, the \emph{Unified Memory} is accessed to retrieve both the synaptic weight and the partial input current of the post-synaptic neuron, enabling both operands to be available for the subsequent execution stage.

	Rather than allocating separate memories for weights and neuron partial currents, SupraSNN employs a unified storage structure. This design choice increases flexibility for the co-design framework, which can trade storage capacity between weights and post-synaptic neuron states depending on the partitioning strategy. To further improve numerical robustness, $K$ weights are packed into a single memory line, while post-synaptic neuron entries store a local neuron index along with a wider partial-current field. Local indices are assigned to internal neurons (excluding input neurons) to reduce index width and memory footprint. Assuming the \emph{Unified Memory} size is $S_{UM}$, the width of \textbf{Post Addr} should be $\log_2 S_{UM}$ to select a single entry from the \emph{Unified Memory}. In contrast, \textbf{Weight Addr} requires an additional $\log_2 K$ bits to select one weight among the $K$ weights packed within a memory line, which is performed using a multiplexer (MUX).
	
	Ideally, the \emph{Unified Memory} would provide two read ports (one for reading the weight and one for the partial current) and a single write port, thereby supporting a throughput of one operation per cycle. To reduce memory area and power consumption, especially in FPGA-based implementations, the \emph{Unified Memory} is instead realized with a single read port, accompanied by a MUX that selects between the \textbf{Weight Addr} and \textbf{Post Addr}. Consequently, each operation is executed over two cycles: the SPU first reads and registers the synaptic weight, and then, in the subsequent cycle, reads the associated partial current. This results in an effective throughput of 0.5 operations per cycle.
	
	\subsubsection{Execution, Write-Back, and ME Packet Generation}
	\noindent In the Execution/Write-Back stage, the synaptic weight is conditionally added to the partial current depending on the spike status of the pre-synaptic neuron. If the pre-synaptic neuron spiked, the updated partial current is written back to \emph{Unified Memory}. When the \textbf{Pre End} flag is asserted, the \emph{Set/Clear Spike} module clears the corresponding spike bit in the associated \emph{Spike Memory} line. The updated line is then written back to the \emph{Spike Memory}, thereby resetting the spike state and preparing the memory for the next timestep.
	
	When the \textbf{Post End} flag is asserted, the partial current accumulated for the post-synaptic neuron represents the final synaptic contribution of that SPU for the current timestep. The partial current is therefore reset by selecting zero, rather than the updated value, through a MUX before the write data port of the \emph{Unified Memory}. Moreover, an ME packet carrying the post-synaptic neuron’s local index and the partial current is injected into the ME Tree.
	
	\section{Centralized Neuron Unit}
	\noindent As discussed previously, neuronal computation in SupraSNN is performed by a centralized Neuron Unit, which time-multiplexes neuron updates and maintains all neuron state. The LIF dynamics are implemented using simple add-and-shift operations derived from \eqref{eq:lif_discrete}--\eqref{eq:vm_reset}, minimizing area and power consumption.
	
\begin{figure}
		\centering
		\includegraphics[width=\linewidth]{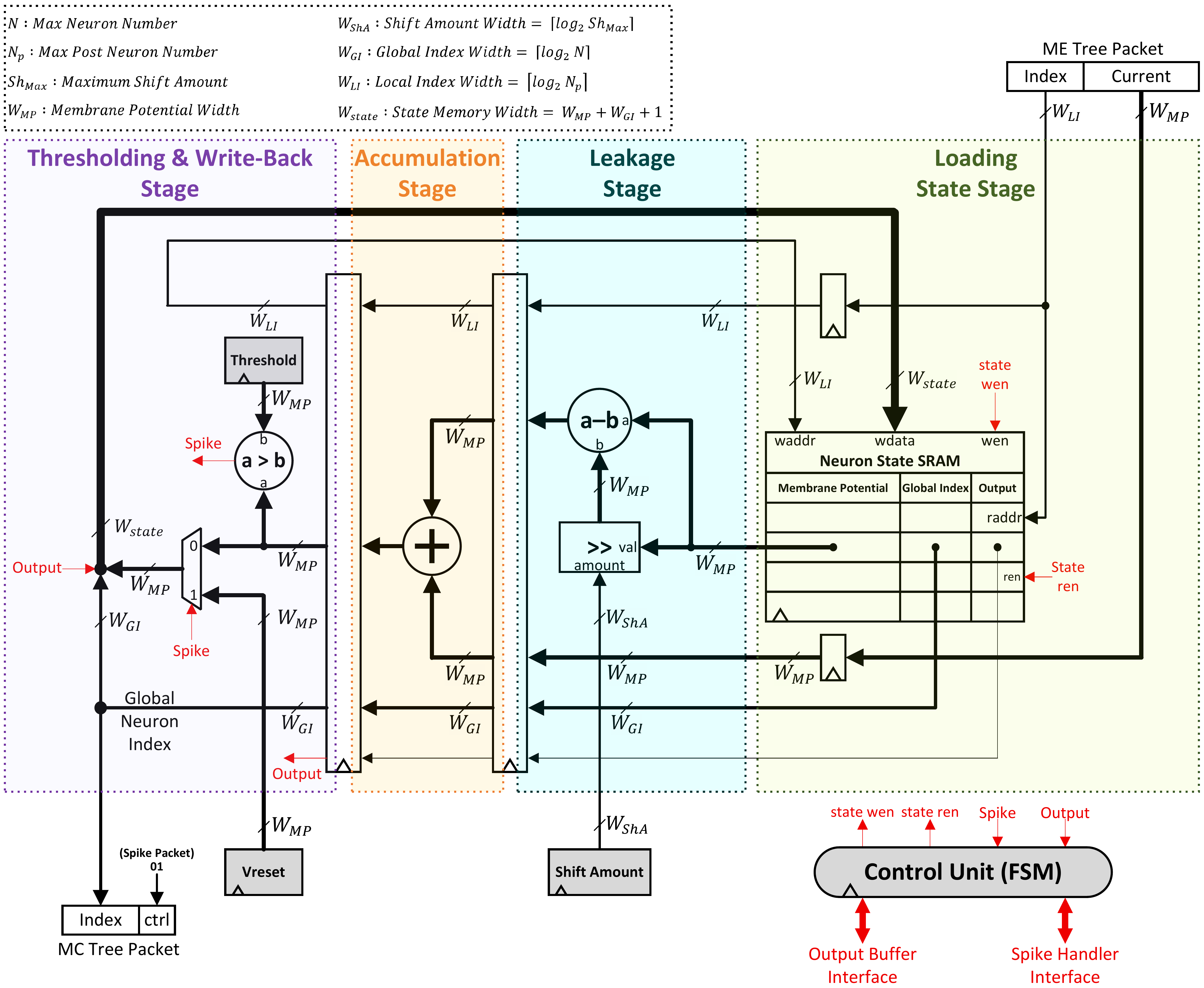}%
		\caption{Micro-architecture of Neuron Unit.}
		\label{fig:Neuron Unit}
	\end{figure}

	Figure~\ref{fig:Neuron Unit} presents the architecture of the Neuron Unit. A programmable \emph{Neuron State SRAM} stores the membrane potential, its global index, and an output flag for each neuron. The global index is necessary because SPUs operate on local indices, while spike packets must reference the neuron's global identifier. The output flag designates neurons whose spike events must be forwarded to both the internal and output buffers. Neuron-model parameters that are not commonly trained online, such as $V_{reset}$, the shift amount implementing $\alpha$, and the firing threshold, are programmed through the hardware bitstream, while the \emph{Neuron State SRAM} is initialized using MC packets explicitly routed to the Neuron Unit, enabling flexible post-fabrication configuration.
	
	The Neuron Unit implements a four-stage pipeline comprising the \emph{Loading State}, \emph{Leakage}, \emph{Accumulation}, and \emph{Thresholding \& Write-Back Stages}. Processing is initiated in the \emph{Loading State Stage} upon the arrival of an ME packet from the ME Tree, which carries the accumulated input current together with the local index of the target post-synaptic neuron. The corresponding neuron-state entry is then fetched from the \emph{Neuron State SRAM}. In the subsequent \emph{Leakage Stage}, the term $(1-\alpha)V_m$ from ~\eqref{eq:lif_discrete} is computed, with the leak factor $\alpha$ approximated to the nearest power of two, replacing multiplications with programmable right-shift operations and reducing hardware complexity. This term is added to the input current in the \emph{Accumulation Stage} to produce $V_m^{updated}$.
	
	During the \emph{Thresholding \& Write-Back Stage}, this value is compared against the firing threshold: if exceeded, a spike-event MC packet containing the neuron's global index is generated and inserted into the Internal Buffer; when the output flag is asserted, the same packet is additionally forwarded to the output buffer. The membrane potential is then reset to $V_{reset}$ upon a spike or updated to $V_m^{updated}$ otherwise, selected through a MUX. Upon receiving an ME packet carrying the \emph{end index}, the Neuron Unit generates a corresponding MC packet and inserts it into both buffers, signaling the Packet Injector that neuronal computation for the current timestep has completed.
	
	\section{Partitioning and Scheduling Framework}
	\subsection{Problem Formulation}
	\noindent SupraSNN exposes parallelism at the synapse level, in contrast to neuromorphic processors that operate primarily at the neuron level. Each SPU contains a \emph{Unified Memory} responsible for storing both synaptic weights and partial post-synaptic currents, while an \emph{Operation Table} stores the mapped synapses and their execution order. We formulate the problem of partitioning synapses among SPUs and scheduling their execution such that memory constraints are satisfied and correctness of the ME Tree merging is preserved.
	
	We model a spiking neural network as a weighted directed graph
	\begin{equation}
	G = (V, E, W),
	\end{equation}
	where $V$ denotes the set of neurons, $E \subseteq V \times V$ denotes the set of directed synapses, and $W : E \rightarrow \mathbb{R}$ assigns a weight to each synapse. Given $M$ SPUs, a partitioning maps each synapse to an SPU via
	\begin{equation}
	\pi : E \rightarrow \{1, \dots, M\},
	\end{equation}
	inducing a synapse cluster
	\begin{equation}
	D_i = \{ e \in E : \pi(e) = i \},
	\end{equation}
	a post-synaptic neuron set $P_i$, and a weight set $Q_i$ for each SPU $i$. Synapse clusters are disjoint ($\forall i \neq j : D_i \cap D_j = \emptyset$),
	whereas post-synaptic neuron sets may overlap ($P_i \cap P_j \neq \emptyset$),
	since synapses targeting the same neuron may be distributed across multiple SPUs, generating partial results that must later be merged.
	
	This partitioning must satisfy the \emph{Unified Memory} capacity constraint on each SPU:
	\begin{equation}
		\forall i : \left\lceil \frac{|Q_i| + 1}{K} \right\rceil + |P_i| \leq L
		\label{eq:UM_constraint}
	\end{equation}
	where the \emph{Unified Memory} contains $L$ memory lines, and each line can store either one post-neuron state or $K$ concatenated synaptic weights. This constraint defines a combinatorial optimization problem that is NP-complete in general, motivating the use of heuristic methods for large-scale SNNs.
	
	While any partitioning that satisfies this constraint is functionally valid, different solutions result in significantly different throughput characteristics. A naïve round-robin partitioning of synapses achieves near-perfect load balance but requires each SPU to store partial currents for almost all post-synaptic neurons, resulting in excessive memory usage. Conversely, grouping all synapses of each post-synaptic neuron onto a single SPU minimizes memory usage but produces severe load imbalance and reduced throughput.
	
	Following partitioning, a scheduling problem determines the execution order of synaptic operations within each SPU. Since SPUs overlap synaptic computation with merging, correctness of the ME Tree requires that the final synapse contributing to a given post-neuron completes in the same cycle across all SPUs storing its partial state. The scheduling therefore augments the partitioned representation with an execution order $\sigma_i$ for each SPU $i$. While any order preserving this alignment is correct, the choice strongly affects throughput: scheduling high fan-in neurons early reduces available slack and increases the critical path, whereas deferring them allows greater overlap between computation and merging. Consequently, scheduling must jointly consider merge-alignment constraints and throughput optimization.
	
\begin{figure}
		\centering
		\includegraphics[width=\linewidth]{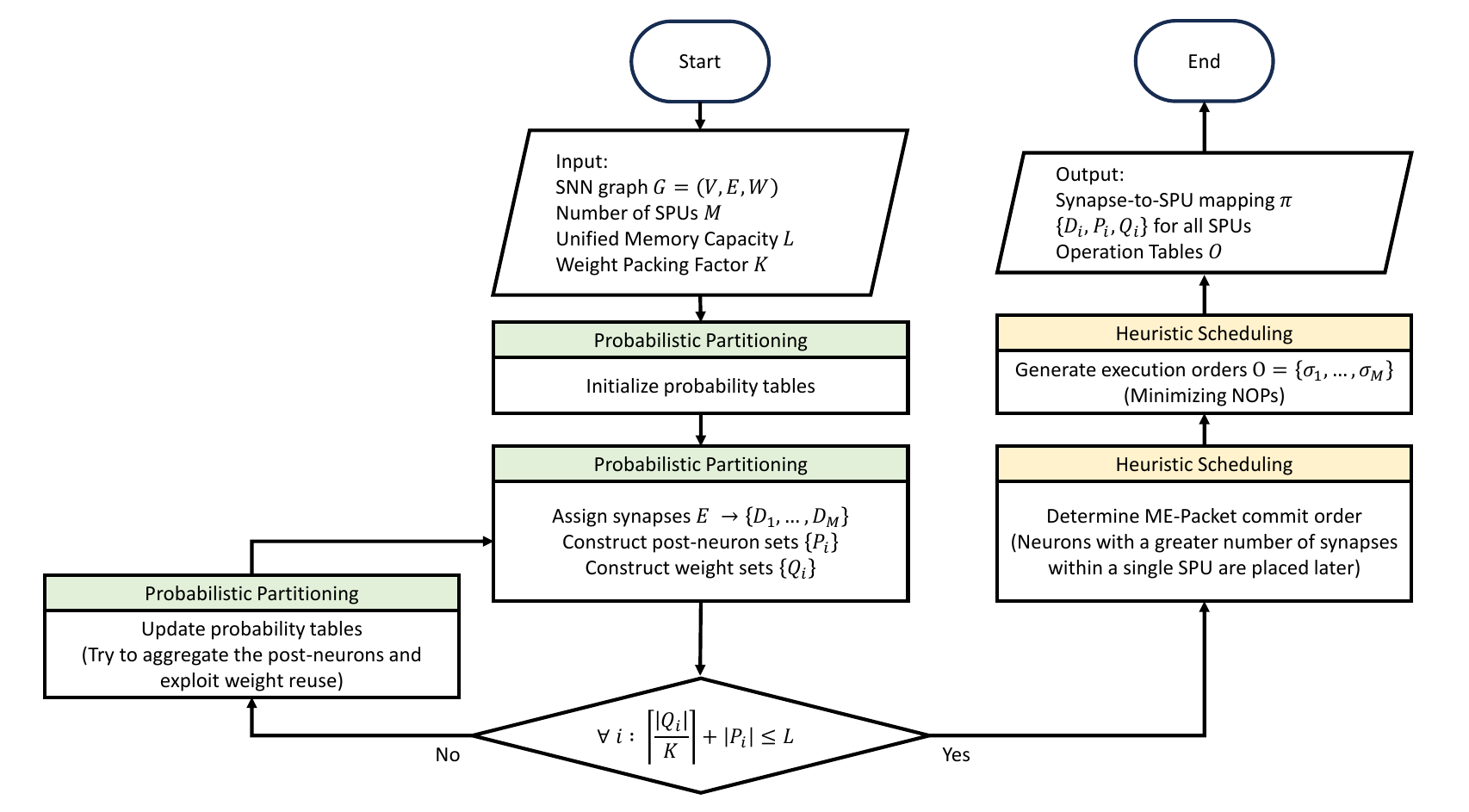}%
		\caption{Flowchart of the software framework responsible for partitioning and scheduling. It has utilized a probabilistic partitioning along with a heuristic scheduling.}
		\label{fig:FlowChart}
	\end{figure}

	To address these challenges, we propose a framework consisting of two components: Probabilistic Partitioning and Heuristic Scheduling, whose overall flow is illustrated in Figure~\ref{fig:FlowChart}.
	
	\subsection{Probabilistic Partitioning Algorithm}
	\noindent In large-scale SNNs, the number of synapses to be mapped onto the multiple SPUs of SupraSNN can be extremely large, making exhaustive optimization impractical due to the exponential complexity of the problem. To address this challenge, we propose a probabilistic partitioning algorithm that efficiently explores the performance--memory trade-off.

\begin{figure}
		\centering
		\includegraphics[width=\linewidth]{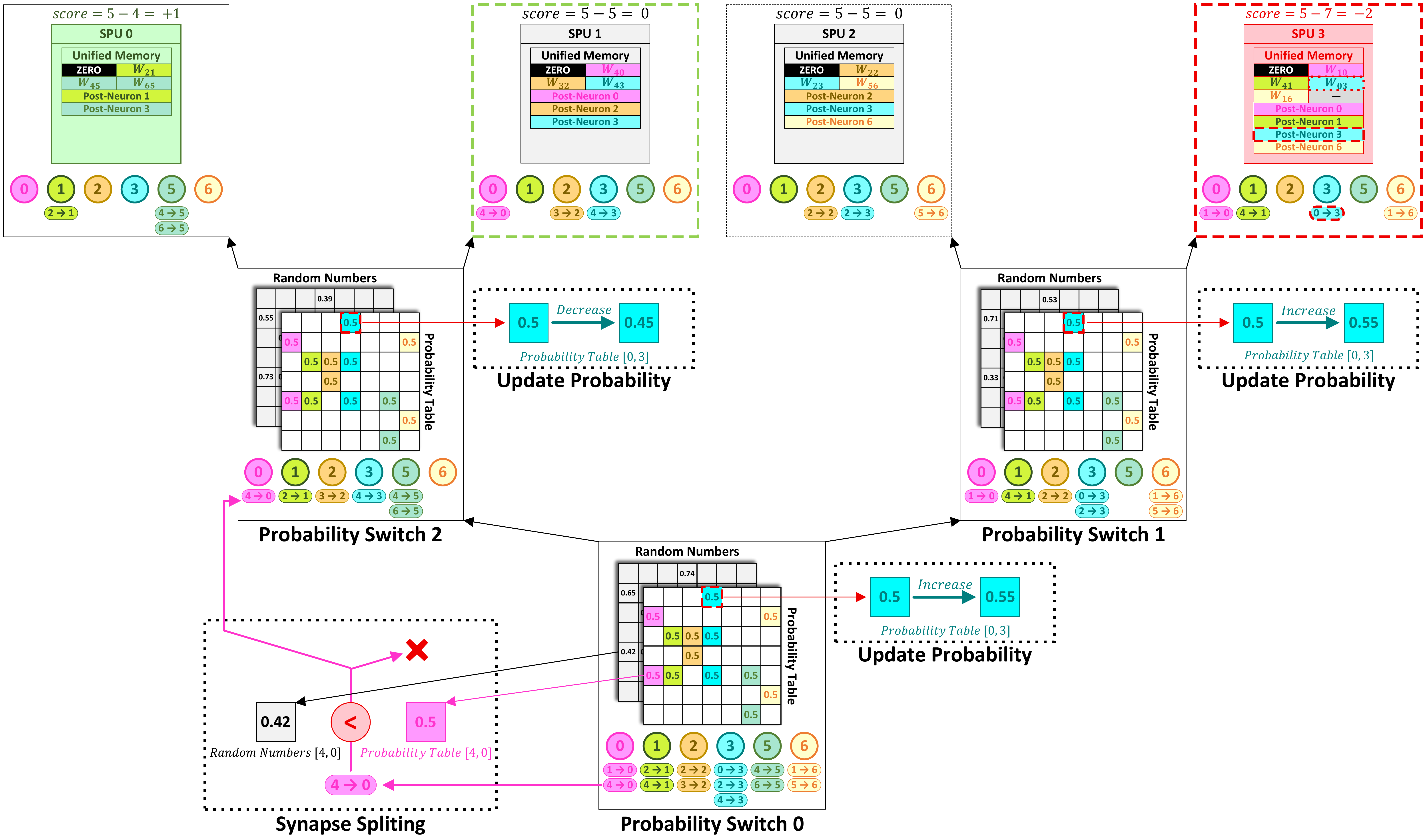}%
		\caption{The overview of the probabilistic partitioning algorithm. The partitioning tree consists of switches with the probability table.}
		\label{fig:Partitioning}
	\end{figure}

	The algorithm employs a binary tree structure, referred to as the \emph{Partitioning Tree}, which mirrors the topology of the ME Tree and is composed of \emph{Probability Switches}. As illustrated in Figure~\ref{fig:Partitioning}, all synapses are initially located at the root, grouped according to their post-neurons and awaiting distribution. Starting from the root, synapses are progressively sent through successive tree levels and \emph{Partitioning Switches} until they reach a leaf SPU, producing a partitioned distribution of the workload. In other words, the switch specifies whether the synapse will ultimately be executed by an SPU located in the left or right subtree. Figure~\ref{fig:Partitioning} illustrates this process for the network from Figure ~\ref{fig:timeline}.
	
	To realize this mechanism, each \emph{Probability Switch} maintains two tables of identical dimensions $|V|\times|V|$, matching the SNN adjacency matrix: a \emph{Probability Table} $P$, where entry $P_{ij}$ represents the probability of sending synapse $i \rightarrow j$ toward the left subtree, and a \emph{Random Numbers Table} $R$, populated with values sampled uniformly from [0,1]. During partitioning, the destination subtree of each synapse is determined by comparing $R_{ij}$ against $P_{ij}$: if $R_{ij} < P_{ij}$, the synapse is sent left; otherwise right. All \emph{Probability Tables} are initialized to 0.5, providing an unbiased starting point that naturally produces a nearly balanced workload distribution across SPUs. The final assignment of synapses to SPUs, shown at the bottom of SPUs in Figure~\ref{fig:Partitioning}, demonstrates that the workload is distributed in a nearly balanced manner. 
	
	After the initial partitioning stage, the memory constraint defined in \eqref{eq:UM_constraint} may be violated by one or more SPUs. To quantify this, a score is assigned to each SPU according to
	\begin{equation}
		\text{Score}_{\text{SPU}_i} = L - \left( \left\lceil \frac{|Q_i| + 1}{K} \right\rceil + |P_i| \right)
		\label{eq:SPU_score}
	\end{equation}
	where $L$ denotes the \emph{Unified Memory} capacity, $Q_i$ represents the set of weights assigned to SPU~$i$, and $P_i$ denotes the set of post-synaptic neurons mapped to that SPU. A negative score indicates a memory violation. In the example shown in Figure~\ref{fig:Partitioning}, the parameters are assumed to be $K = 2$ and $L = 5$. Under this configuration, SPU~0 achieves the highest score of $+1$, while SPU~1 and SPU~2 satisfy the memory constraint with a score of $0$. In contrast, SPU~3 exceeds the memory capacity $L = 5$, resulting in the lowest score of $-2$. The SPU with the minimum score is therefore identified as the overloaded unit (SPU~3), whereas SPUs with higher scores are candidates for receiving redistributed workload.
	
	To alleviate the memory violation, a synapse is selected from the overloaded SPU for reassignment. Preference is given to synapses whose post-neuron is not shared with any other synapse within the same SPU (i.e., the last synapse associated with that post-neuron in the SPU). Removing such a synapse eliminates both its corresponding weight and the associated post-neuron entry, thereby freeing an entire memory line plus an additional $1/K$ fraction of memory. For instance, if a synapse were to be selected from SPU~0, synapse $4 \rightarrow 5$ would not be preferred because the SPU also contains synapse $6 \rightarrow 5$. In SPU~3, which is the most overloaded SPU in the example, synapse $0 \rightarrow 3$ is selected for reassignment.
	
	The algorithm then identifies the most suitable destination SPU among the higher-scored candidates using the following priority order:
	\begin{enumerate}
	\item the highest-scored SPU that already contains both the corresponding post-neuron and weight,
	\item the highest-scored SPU that contains the post-neuron,
	\item the highest-scored SPU that contains the weight, and
	\item otherwise, the highest-scored SPU.
	\end{enumerate}
	This priority order exploits post-neuron sharing and weight reuse to minimize the additional memory overhead incurred by the reassignment, with post-neuron sharing given higher priority since storing a post-neuron state occupies a full memory line compared to only $1/K$ of a line for a weight. In our example, while SPU~0 has the highest score, SPU~1 and 2 are more favorable choices since they have already stored post-neuron 3 entry. Introducing an assumption that $W_{03} = W_{32}$, SPU~1 incurs no additional memory overhead and is identified as the most suitable destination for synapse $0 \rightarrow 3$.
	
	To enforce the reassignment probabilistically, the partitioning tree is traversed from both the overloaded SPU (SPU~3) and the selected destination SPU (SPU~1) toward their lowest common ancestor (\emph{Probability Switch~0}). Along the path from the overloaded SPU, probability entries are adjusted to decrease the likelihood that the selected synapse is sent to that subtree (e.g. $P_{03}$ is increased in \emph{Probability Switch~1}). Conversely, the probabilities along the path toward the underloaded SPU are modified to increase the likelihood that the synapse is routed to that subtree. The partitioning process is then repeated using the updated tables, and this iterative procedure continues until all SPUs satisfy the memory constraint or a predefined iteration limit is reached.
	
	A key design consideration concerns whether the \emph{Random Numbers Tables} should be regenerated at each iteration or kept fixed. Regenerating them increases exploration but makes probability updates ineffective, preventing convergence. Keeping them fixed preserves the feedback mechanism but risks convergence to a local minimum. We adopt the latter approach and incorporate a perturbation mechanism to escape local minima: the framework monitors the average SPU scores over the previous 100 iterations, and if this value fluctuates within a range smaller than 0.2 -- indicating stagnation -- a uniformly distributed random value in [-0.1, 0.1] is added to each entry of the \emph{Random Numbers Tables}. This controlled perturbation slightly alters the partitioning behavior, allowing the search to escape the local minimum and continue progressing toward a feasible solution.      
	
	Overall, the algorithm starts from a balanced initial partitioning and progressively clusters synapses sharing post-neurons and weights to reduce memory usage, at the necessary cost of more unbalanced synapse distribution and lower throughput. The use of probability tables within the partitioning tree enables smooth exploration of the design space while converging toward a feasible partitioning.
	
	\subsection{Heuristic Scheduling Algorithm}
	\noindent After the synapses are assigned to the SPUs, their execution order must be scheduled. The key constraint in this scheduling problem is that all SPUs must inject the ME packets associated with the same post-neuron into the ME Tree during the same clock cycle. This requirement ensures the correctness of the merging operation performed by the tree. Subject to this constraint, the objective of the scheduler is to minimize overall latency.
	
\begin{figure}
		\centering
		\includegraphics[width=\linewidth]{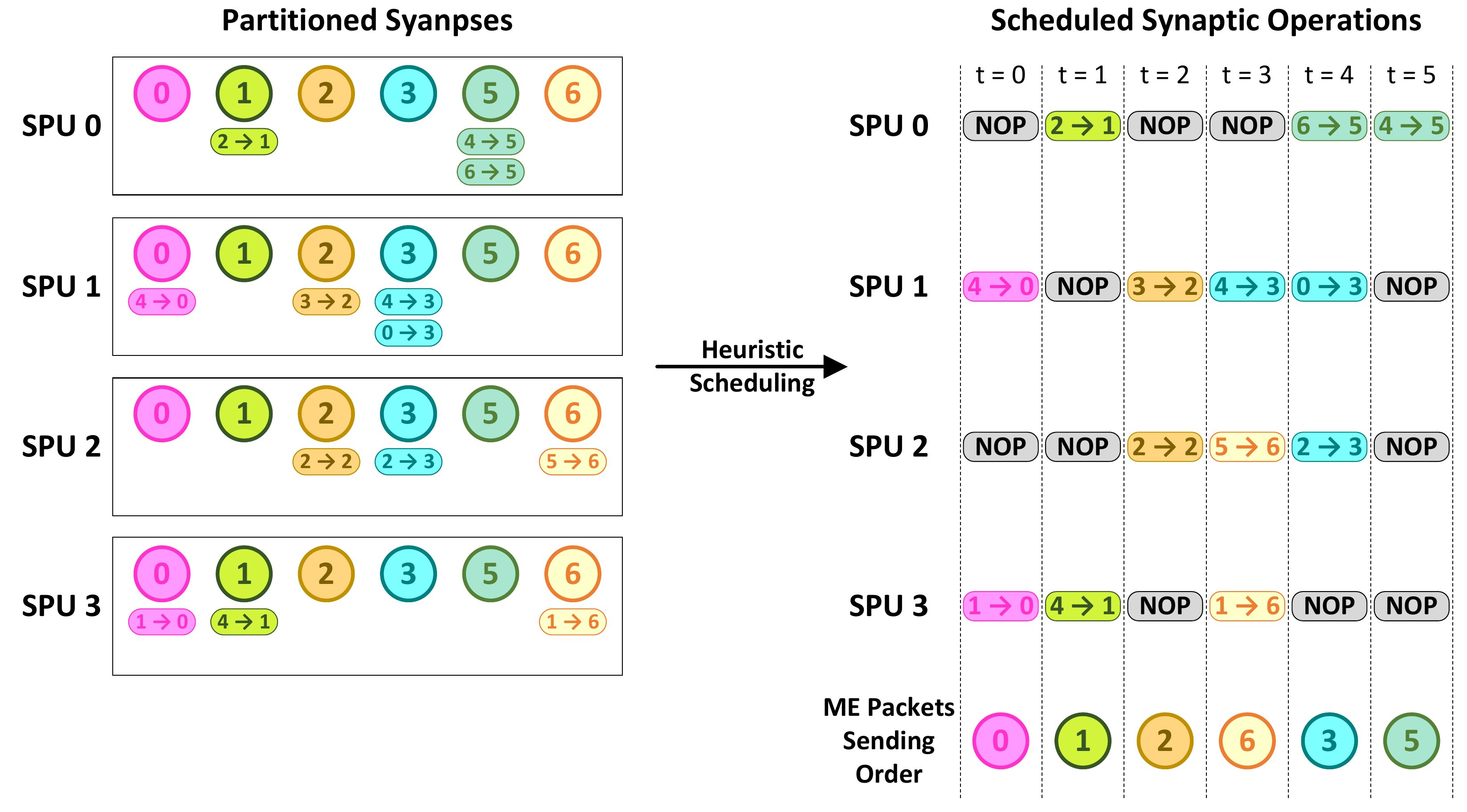}%
		\caption{The overview of the heuristic scheduling algorithm.}
		\label{fig:Scheduling}
	\end{figure}

	Figure~\ref{fig:Scheduling} illustrates the proposed scheduling strategy using the sample SNN depicted in the top-right corner of Figure~\ref{fig:timeline}. The left side of the figure shows the synapse partitions assigned to each SPU, organized according to their post-neurons.
	
	To schedule the assigned synapses while satisfying the synchronization constraint, the algorithm first determines the ME-packet sending order. To maximize the available
	slack for synaptic computations, post-neurons are sorted in ascending order based on the maximum number of synapses they have on any single SPU. This ordering delays the injection of post-neurons that require more synaptic operations on a particular SPU, giving that SPU additional cycles to complete its computations before transmission.
	In the example, Neuron~5 has two synapses on SPU~0, and Neuron~3 has two synapses on SPU~1. Consequently, these post-neurons are placed at the end of the sending sequence. All other post-neurons contain at most one synapse per SPU and therefore appear earlier. This heuristic produces the ME-packet sending order shown at the bottom of Figure~\ref{fig:Scheduling}: 0, 1, 2, 6, 3, and 5.
	
	Once the sending order is determined, the execution order of synapses is specified. As described in the SupraSNN architecture, each SPU transmits the ME packet of a post-neuron immediately after its last synapse is computed. Accordingly, one synapse of each post-neuron is scheduled at its corresponding sending time. If a post-neuron has no synapse in a given SPU, the corresponding time slot remains empty and can later be utilized by other synapses.
	The remaining synapses are scheduled backward in time, starting from the last post-neuron in the sending order. The algorithm traverses cycle by cycle in reverse, placing synapses in available empty slots. This process is repeated sequentially for each post-neuron in reverse sending order.
	
	For example, Neuron~5 has two synapses on SPU~0. Its final synapse (4~$\rightarrow$~5) is placed at $t = 5$, the ME-packet time for Neuron~5. To schedule the second synapse (6~$\rightarrow$~5), the scheduler searches backward from $t = 5$ and finds that $t = 4$ is empty on SPU~0 because that SPU contains no synapse for Neuron~3, which transmits at $t = 4$. Starting from the last post-neuron is logical because its final synapse has already been fixed at the sending time, and no further synapses of that neuron may be scheduled afterward.
	
	Finally, any remaining empty slots are filled with NOP operations to maintain cycle-level synchronization across all SPUs.
	
	\section{Experimental Evaluation}
	\subsection{Baseline Results}
	\noindent SupraSNN addresses digit classification in two modalities: visual and auditory digit recognition. For evaluation, we employ two standard benchmark datasets: MNIST \cite{MNIST} for visual digits and SHD \cite{SHD} for audio-based spike representations of spoken digits.
	
	MNIST is a widely used baseline for assessing AI hardware implementations, which makes it an ideal choice for comparing our design with existing accelerators. This dataset consists of 60,000 training and 10,000 test samples, each being a $28 \times 28$ grayscale image of handwritten digits from 0 to 9.
	
	On the other hand, SHD is a dataset specifically designed for SNN applications. It contains approximately 10,000 high‑quality recordings from 12 speakers pronouncing the digits 0 to 9 in both English and German. These recordings are converted into spike trains using an artificial model of the inner ear and parts of the ascending auditory pathway. This dataset enables us to explore more complex network topologies, such as recurrent structures.
	
	 For the MNIST dataset, we trained a SNN with the architecture of 784 input neurons, corresponding to the $28 \times 28$ pixels of the MNIST images, a hidden layer with 116 neurons, and 10 output neurons. This network follows a purely feedforward architecture and does not include recurrent connections.
	
	To deploy these datasets on SupraSNN, the parameters of the LIF neuron model are first configured as summarized in Table~\ref{tab:parameters}. For MNIST, rate encoding is used to convert each input pixel into a spike train proportional to its intensity, whereas SHD already provides spike-based input data. In both networks, classification is performed at the output layer by selecting the neuron with the highest accumulated spiking activity.
	
	The networks are implemented using snnTorch \cite{snnTorch} and trained with the Back-Propagation Through Time (BPTT) algorithm. The surrogate gradient functions and other training hyperparameters are summarized in Table~\ref{tab:parameters}. After training, the models achieve classification accuracies of 96.30\% on MNIST and 71.02\% on SHD. To improve computational efficiency and better match the target hardware architecture, synaptic pruning is applied during training. Starting from an initially fully connected network containing 92,604 synapses for MNIST and 306,000 synapses for SHD, binary sparsity masks are used to remove a portion of connections before training. This process results in final sparsity levels of 51.89\% for MNIST and 87.04\% for SHD.
	
	\begin{table}
		\centering
		\caption{Model, Training, Hardware and Implementation Results}
		\label{tab:parameters}
		\footnotesize
		\resizebox{0.6\linewidth}{!}{
		\begin{tabular}{llcc}
			\toprule
			& Parameter & MNIST & SHD \\
			\midrule
			
			\multirow{7}{*}{Model}
			& $\alpha$ & 0.25 & 0.03125 \\
			& $V_{reset}$ & 0.0 & 0.0 \\
			& $V_{threshold}$ & 1.0 & 1.0 \\
			& Architecture & 784-116-10 & 700-300-20 \\
			& Timestep & 10 & 100 \\
			& Encoding & Rate Code & Custom \\
			& Network Type & SFNN & SRNN \\
			
			\midrule
			\multirow{7}{*}{Training}
			& Method & BPTT & BPTT \\
			& Surrogate Function & ReLU & Sigmoid \\
			& Learning Rate & $5\times10^{-4}$ & $10^{-5}$ \\
			& Optimizer & Adam & Adam \\
			& Epoch & 20 & 60 \\
			& Sparsity & 51.89\% & 87.04\% \\
			& Accuracy & 96.30\% & 71.02\% \\
			
			\midrule
			\multirow{9}{*}{Hardware}
			& SPU & 16 & 64 \\
			& Unified Mem Depth & 128 & 256 \\
			& Operation Table & 661 & 742 \\
			& Weight Width & 4 & 7 \\
			& Concentration Factor ($K$) & 3 & 3 \\
			& Potential Width & 5 & 12 \\
			& Max Neurons & 910 & 1020 \\
			& Max Post-Neurons & 126 & 320 \\
			& Clock Frequency (MHz) & 100 & 100 \\
			
			\midrule
			\multirow{13}{*}{Result}
			& FPGA & XC7Z020 & XC7Z030 \\
			& Avail LUT & 53,200 & 78,600 \\
			& Used LUT & 3,092 (5.81\%) & 14,995 (19.08\%) \\
			& Avail FF & 106,400 & 157,200 \\
			& Used FF & 3,052 (2.87\%) & 14,945 (9.51\%) \\
			& Avail BRAM & 140 & 265 \\
			& Used BRAM & 33.5 (23.92\%) & 131 (49.43\%) \\
			& Post-Quantization Sparsity & 88.74\% & 88.19\% \\
			& Static Power (W) & 0.106 & 0.130 \\
			& Dynamic Power (W) & 0.066 & 0.416 \\
			& Total Power (W) & 0.172 & 0.546 \\
			& Latency (ms) & 0.149 & 1.41 \\
			& Accuracy & 93.44\% & 71.82\% \\
			
			\bottomrule
		\end{tabular}
		}
	\end{table}
	
	To deploy the trained networks on SupraSNN, various hardware configurations are explored to achieve a suitable balance between performance and resource efficiency. The final hardware parameters selected for implementation are summarized in Table~\ref{tab:parameters}. Due to the larger network size, higher complexity of the SHD dataset, and larger number of timesteps, a wider numerical representation and a greater number of parallel SPUs are allocated compared with the MNIST configuration.
	
	The finalized hardware configurations for the MNIST and SHD networks are implemented on Xilinx Zynq XC7Z020 and XC7Z030 FPGAs, respectively, using the Vivado design suite. Resource utilization and power consumption are obtained through the built-in analysis tools provided by Vivado, summarized in Table~\ref{tab:parameters}. The XC7Z030 device is selected for the SHD implementation due to its larger available hardware resources, which accommodate the more demanding configuration required by the SHD network.
	
	\subsection{Comparison with Other Accelerators}
	\noindent Table~\ref{tab:comparison_table} summarizes the results reported by state‑of‑the‑art FPGA‑based SNN accelerators that use MNIST as their evaluation benchmark. It reports the results of SNN accelerators designed for fully connected network topologies. The accuracy achieved by our implementation (93.44\%) demonstrates competitive performance and remains comparable with other FPGA‑based accelerators evaluated on the same benchmark.
	
	\begin{table}
		\caption{Comparison of SupraSNN to FPGA implementations of SNN accelerators}
		\centering
		\resizebox{\linewidth}{!}{
			\begin{tabular}{lcccccc}
				\toprule
				Design & Han et al. \cite{Han et al.} & Gupta et al. \cite{Gupta et al.} & Li et al. \cite{Li et al.} & Spiker \cite{Spiker} & Spiker+ \cite{Spiker+} & SupraSNN (This Work) \\
				\midrule
				Year & 2020 & 2020 & 2021 & 2022 & 2024 & 2026 \\
				Clk Frequency (MHz) & 200 & 100 & 100 & 100 & 100 & 100 \\
				Timestep & N/R & N/R & N/R & 3500 & 100 & 10 \\
				Potential Width & 16 & 24 & 16 & 16 & 6 & 5 \\
				Weight Width & 16 & 24 & 16 & 16 & 4 & 4 \\
				Update & Event & Event & Hybrid & Clock & Clock & Clock \\
				Model & LIF & LIF & LIF & LIF & LIF & LIF \\
				FPGA & XC7Z045 & XC6VLX240T & XC7VX485 & XC7Z020 & XC7Z020 & XC7Z020 \\
				Avail BRAM & 545 & 416 & 2060 & 140 & 140 & 140 \\
				Used BRAM & 40.5 & 162 & N/R & 45 & 18 & 33.5 \\
				Avail DSP & 900 & 768 & 2800 & 220 & 220 & 220 \\
				Used DSP & 0 & 64 & N/R & 0 & 0 & 0 \\
				Avail Logic Cells & 655800 & 452160 & 485760 & 159600 & 159600 & 159600 \\
				Used Logic Cells & 12690 & 79468 & N/R & 55998 & 7612 & 6144 \\
				Architecture & 784-1024-1024-10 & 784-16 & 784-200-100-10 & 784-400 & 784-128-10 & 784-116-10 \\
				Synapses & 1861632 & 12544 & 177800 & 313600 & 101632 & 92604 \\
				Latency/Image (ms) & 6.21 & 0.50 & 3.15 & 0.22 & 0.78 & 0.149 \\
				Power (W) & 0.477 & N/R & 1.6 & 59.09 & 0.18 & 0.172 \\
				Energy/Image (mJ) & 2.96 & N/R & 5.04 & 13 & 0.14 & 0.02563 \\
				Energy/Synapse (nJ) & 1.59 & N/R & 28 & 41 & 1.37 & 0.27675 \\
				Accuracy & 97.06\% & N/R & 92.93\% & 73.96\% & 93.85\% & 93.44\% \\
				\bottomrule
			\end{tabular}
		}
		\label{tab:comparison_table}
	\end{table}
	
	In terms of latency, SupraSNN achieves the best performance, classifying a single MNIST image with an average latency of 0.149 ms, which is 47.6\% lower than the second‑best design reported by Spiker \cite{Spiker}. This performance advantage stems from two key architectural characteristics. First, SupraSNN achieves high utilization of its parallel processing units by exploiting \emph{synapse‑level parallelism}. Second, it effectively leverages weight sparsity by avoiding the storage and execution of synapses with zero weights.
	
	The importance of these design choices can be illustrated by comparing SupraSNN with Spiker+ \cite{Spiker+}. Spiker+ achieves a high level of hardware parallelism by assigning a dedicated processing unit to each neuron, resulting in 138 processing units, which is approximately nine times larger than the number of SPUs used in SupraSNN. However, this design leads to imbalanced workload distribution across processing units. In particular, processing units associated with the first layer are responsible for 784 synapses, whereas those in the output layer process only 128 synapses per timestep. This imbalance, combined with the execution of zero‑valued weights and the use of larger simulation timesteps, results in a significantly higher inference latency—approximately $5.2 \times$ greater than that of SupraSNN.
	
	SupraSNN requires 33.5 BRAMs (23.93\% of the device capacity), making it the second most memory‑efficient design after Spiker+. Although \emph{synapse‑level parallelism} introduces additional memory overhead to store synaptic metadata in the \emph{Operation Table}, the architecture simultaneously benefits from not storing zero‑valued synapses, which effectively mitigates this overhead. Consequently, SupraSNN maintains a small memory footprint, particularly when executing sparse networks.
	
	For logic resources, we report utilization as the sum of LUTs and flip‑flops, which are 3,092 LUTs (5.81\% of the available) and 3,052 FFs (2.87\% of the available) in our implementation. SupraSNN achieves the lowest logic utilization among all pure spiking accelerators listed in Table~\ref{tab:comparison_table}, reflecting its compact hardware footprint. The next most area‑efficient design is Spiker+, which requires 7,612 logic cells on the same FPGA device, representing a 23.9\% increase compared to SupraSNN.
	
	SupraSNN also exhibits the lowest power consumption among the fully connected accelerators considered in Table~\ref{tab:comparison_table}. Although the difference relative to Spiker+ is modest (approximately 0.008 W), the combination of lower power and the shortest latency enables SupraSNN to achieve the best energy efficiency per image by a substantial margin. The second‑best design, Spiker+, requires approximately $5.6 \times$ more energy per inference.
	
	To account for the varying network sizes across different accelerators, we also report a normalized energy‑per‑synapse metric. SupraSNN consumes only 0.27675 nJ per synapse on average to classify a single MNIST image, demonstrating its high efficiency in utilizing available synaptic operations.
	
	\subsection{Weight Sparsity Evaluation in SupraSNN}
	\noindent In this section, we aim to study how effectively SupraSNN exploits weight sparsity regarding performance, memory, power, and energy. To this end, we trained a network using the parameters and methods mentioned in Table~\ref{tab:parameters} on the SHD dataset across various sparsity levels.
	
\begin{figure}
		\centering
		\includegraphics[width=\linewidth]{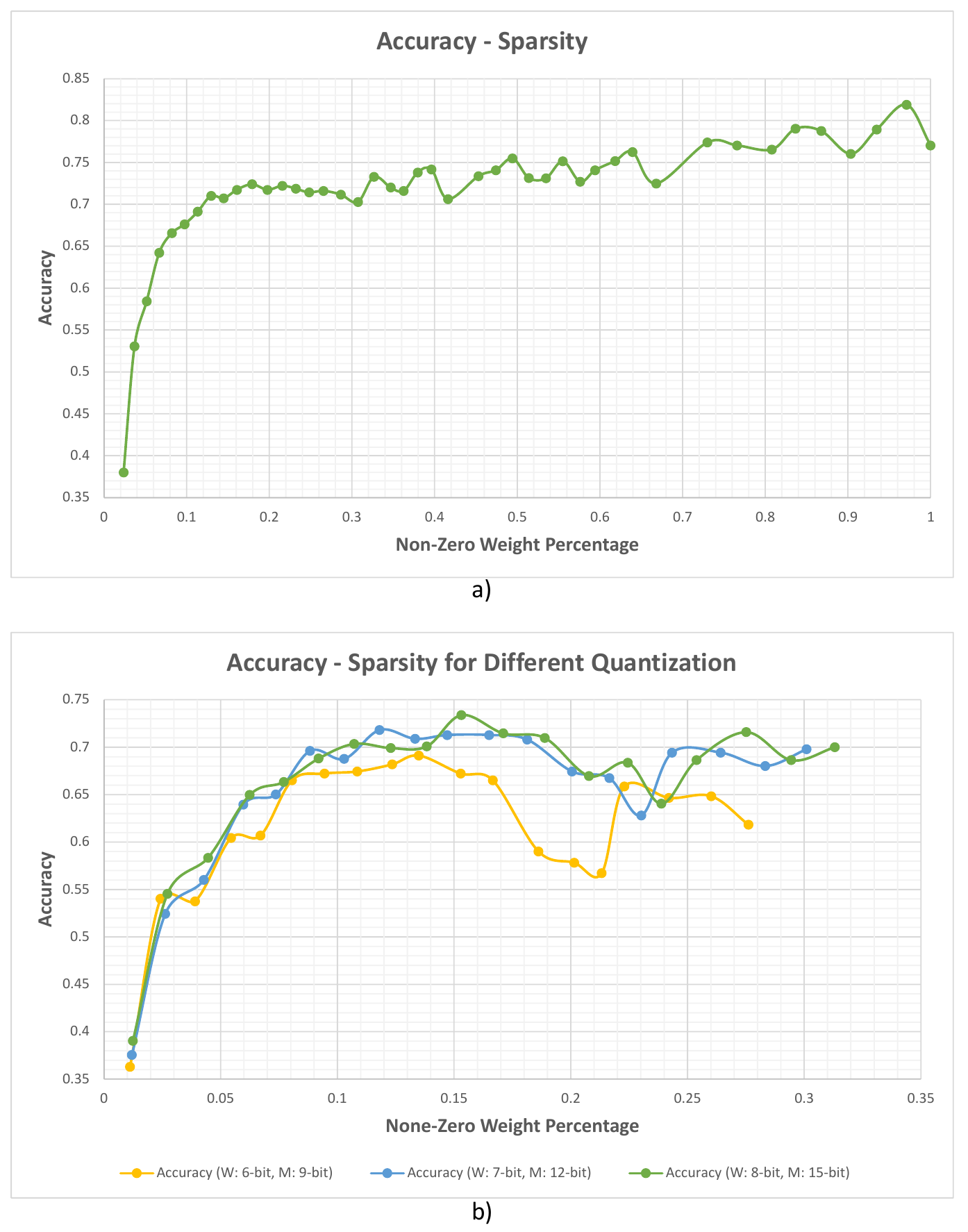}%
		\caption{(a) Accuracy of the 700–300–20 SNN architecture at different sparsity levels. (b) Accuracy of the quantized SNN for three quantization bit widths across varying sparsity levels.}
		\label{fig:sparsity accuracy}
	\end{figure}

	Figure~\ref{fig:sparsity accuracy}a shows the software accuracy across all 2,264 SHD test samples for various sparsity levels. While the peak accuracy of 81.89\% occurs at low sparsity (dense connectivity), networks with substantially fewer non-zero synapses still achieve strong performance. Notably, a network with 82.1\% sparsity acquired 72.39\% accuracy, representing the elbow point of the curve.
	
	Figure~\ref{fig:sparsity accuracy}b compares three quantization configurations in terms of hardware accuracy. The configuration with 8-bit weights and 15-bit membrane potential generally achieves the highest accuracy, though performance differences become less pronounced as sparsity increases. Consequently, the configuration with 6-bit weights and 9-bit membrane potential represents a more practical choice for sparsity evaluation, as it significantly reduces memory requirements for weights, synaptic currents, and membrane potentials while maintaining acceptable accuracy at high sparsity levels.
	
\begin{figure}
		\centering
		\includegraphics[width=\linewidth]{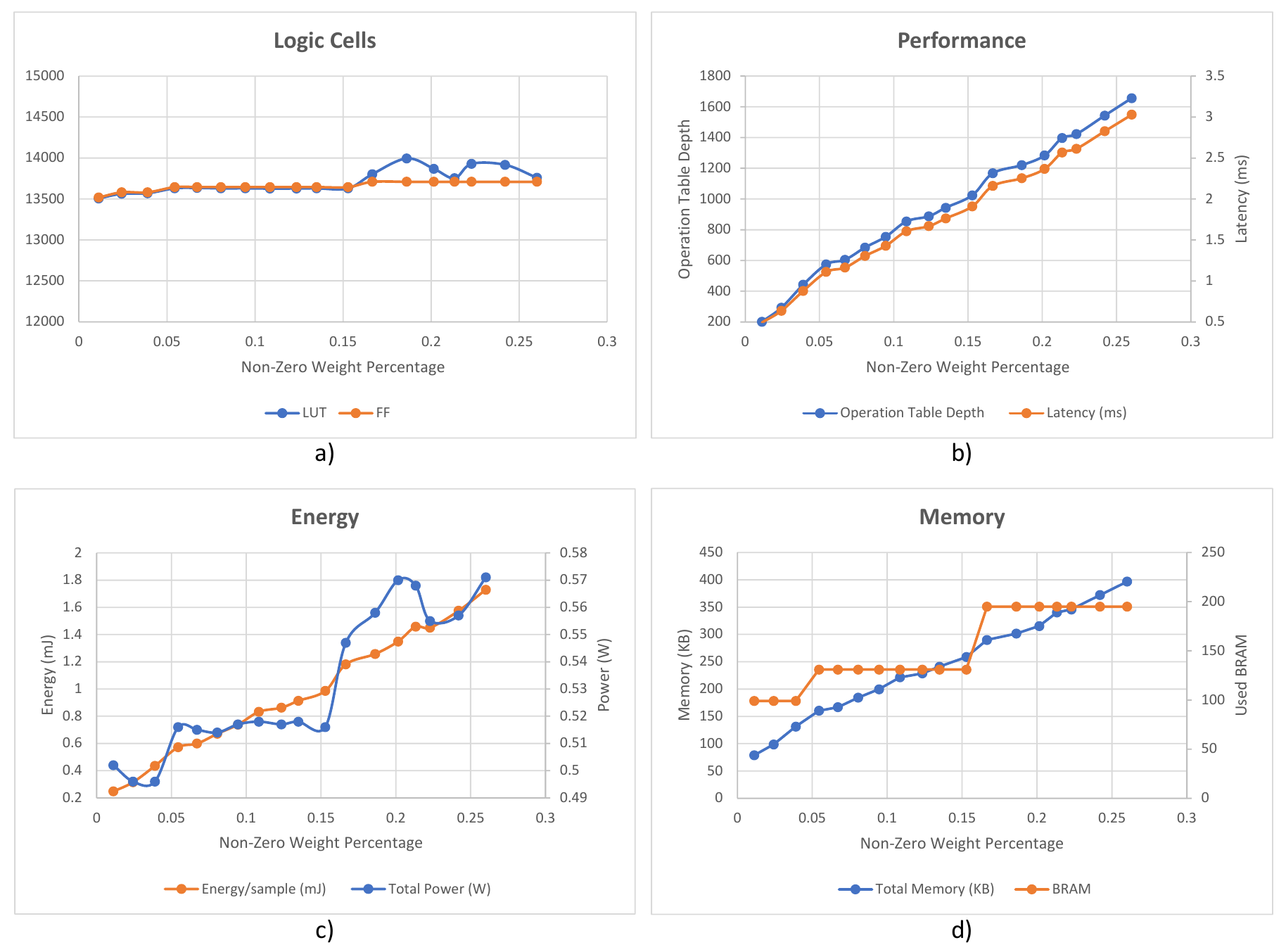}%
		\caption{(a) Total number of LUTs and FFs utilized by the trained SNN at different sparsity levels (b) \emph{Operation Table} depth and the average inference latency (in ms) based on the sparsity level (c) Total power on the FPGA and average energy per inference across different sparsity levels (d) Memory footprint in KB and the BRAM utilization for different sparsity levels.}
		\label{fig:sparsity_evaluation}
	\end{figure}

	For each of the 18 trained SHD networks, the model is first processed by our Partitioning and Scheduling framework, which partitions synapses across the SPUs, schedules the execution sequence, determines a feasible mapping, identifies the minimum required \emph{Operation Table} depth, and generates the initialization packets and bitstream. The framework also estimates the average classification latency per sample through cycle‑accurate simulation. Hardware resource utilization and power consumption is obtained by synthesizing the generated designs in Vivado targeting the XC7Z030 Zynq FPGA. The results are presented in Figure~\ref{fig:sparsity_evaluation}.
	
	Figure~\ref{fig:sparsity_evaluation}a shows that both LUT and FF utilization remain relatively constant across all sparsity levels, indicating that logic resource usage is primarily determined by architectural parameters such as the number of SPUs, rather than by the number of synaptic connections in the mapped network.
	
	In contrast, the required \emph{Operation Tables} depth grows directly with the number of non zero synapses, as shown in Figure~\ref{fig:sparsity_evaluation}b, since only non-zero synapses are stored in the SPU memory units. Denser networks therefore require deeper \emph{Operation Tables} and longer execution sequences. A 24.89\% increase in the proportion of non-zero synapses leads to a 2.5352 ms increase in latency, highlighting the effectiveness of the design in exploiting weight sparsity to achieve significant latency reductions in sparse networks.
	
	Figure~\ref{fig:sparsity_evaluation}c shows that the measured FPGA power does not follow a perfectly monotonic trend, it generally increases as networks become denser.
	
	Combined with the latency trend observed in Figure~\ref{fig:sparsity_evaluation}-b, the per-sample classification energy consumption increases accordingly, since energy is proportional to the product of power and execution time. Conversely, sparser SNNs benefit from shorter execution latency and slightly lower power, leading to reduced energy per classified sample.
	
	To further evaluate how SupraSNN exploits weight sparsity, the overall memory footprint of the architecture must be analyzed as a function of the design parameters. Let the number of SPUs be $M$, the maximum number of neurons be $N$, the maximum number of post neurons be $N_{p}$, the weight bit width be $W_{W}$, the weight concatenation factor be $K$, the Operation Table depth be $S_{OT}$, and the \emph{Unified Memory} depth be $S_{UM}$.
	
	Three components of the architecture contain memory structures: (1) the Routing Unit, (2) the SPUs, and (3) the Neuron Unit.
	
	The Routing Unit stores a routing bitstring for each neuron, where the bitstring width equals the number of SPUs ($M$). Therefore, the total routing memory capacity is $N \times M$ bits.
	
	Each SPU contains two memory structures: the \emph{Operation Table} and the \emph{Unified Memory}. Every operation entry in the \emph{Operation Table} stores several fields:
	
	\begin{itemize}
		\item a Post Addr field requiring $\lceil \log_{2} S_{UM} \rceil$ bits,
		\item a Weight Addr field requiring $\lceil \log_{2} S_{UM} \rceil + \lceil \log_{2} K \rceil$ bits,
		\item a Spike Addr field requiring $\lceil \log_{2} N \rceil$ bits, and
		\item two control bits indicating the Pre End and Post End conditions.
	\end{itemize}
	
	Accordingly, the bit width of each Operation Table entry becomes:
	\[
	2\lceil \log_{2} S_{UM} \rceil + \lceil \log_{2} K \rceil + \lceil \log_{2} N \rceil + 2
	\]
	
	Given an \emph{Operation Table} depth of $S_{OT}$, the total capacity of the \emph{Operation Table} in each SPU is:
	\[
	S_{OT} \times \left(2\lceil \log_{2} S_{UM} \rceil + \lceil \log_{2} K \rceil + \lceil \log_{2} N \rceil + 2\right)
	\]
	
	The \emph{Unified Memory} stores concatenated weights. Its width is $K \times W_W$ bits and its depth is $S_{UM}$, leading to a total capacity of:
	\[
	K \times W \times S_{UM}
	\]
	
	The Neuron Unit also includes a memory structure for storing neuron states. Each neuron state contains:
	
	\begin{itemize}
		\item $1$ bit indicating whether the neuron is an output neuron,
		\item $\lceil \log_{2} N \rceil$ bits for the neuron’s global index, and
		\item $KW - \lceil \log_{2} N_p \rceil$ bits for the membrane potential.
	\end{itemize}
	
	Since the neuron state memory has a depth of $N_p$, its total capacity becomes:
	\[
	N_p \times \left(\lceil \log_{2} N \rceil + KW_W - \lceil \log_{2} N_p \rceil + 1 \right)
	\]
	
	By combining all memory components, the total memory requirement of the proposed architecture can be expressed as:
	\begin{equation}
		\label{eq:required_memory}
		\begin{aligned}
			Mem_{total} =\ &NM 
			+ M \Big( S_{OT}\big( 2\lceil\log_{2} S_{UM}\rceil 
			+ \lceil\log_{2} K\rceil \\
			&+ \lceil\log_{2} N\rceil 
			+ 2 \big)
			+\ KW_W S_{UM} \Big) \\
			&+ N_p\Big( \lceil\log_{2} N\rceil 
			+ KW_W 
			- \lceil\log_{2} N_p\rceil 
			+ 1 \Big)
		\end{aligned}
	\end{equation}
	
	Figure~\ref{fig:sparsity_evaluation}d reports the total memory footprint and BRAM utilization across all evaluated networks. As networks become denser, the required memory increases from 78.68 KB to 396.96 KB -- more than a fivefold increase -- while the proportion of non-zero synapses rises by only 24.89\%. BRAM utilization increases in three distinct steps as network density grows. This strong dependence of memory footprint on sparsity confirms that SupraSNN effectively reduces memory requirements for sparse networks, providing an opportunity to map larger SNN models onto resource-constrained FPGA platforms.
	
	Overall, the results demonstrate that SupraSNN dynamically scales its latency, power, and memory footprint in direct proportion to network sparsity, while logic resources remain unaffected. This confirms the effectiveness of the \emph{operation-based} execution model in translating algorithmic sparsity into tangible hardware efficiency gains across all key metrics.
	
	\subsection{Partitioning and Scheduling Evaluation in SupraSNN}
	\noindent To evaluate the effectiveness of the proposed partitioning and scheduling framework, we utilize an SNN configuration with 9-bit weight width and 18-bit membrane potential width. A larger weight precision is intentionally chosen to better demonstrate the framework's weight reuse capability: with smaller widths such as 4 bits, the number of distinct weight values is severely limited, making weight storage overhead negligible compared to neuron state storage and obscuring the benefits of weight reuse. After quantization, the resulting network contains 33,457 non-zero synapses, achieves 70.71\% classification accuracy on hardware, and includes 289 unique weight values.
	
	To evaluate the proposed framework, we execute it under different \emph{Unified Memory} depth constraints, which represent the primary resource limitation of the mapping problem. In total, 48 configurations are examined, with \emph{Unified Memory} depths ranging from 46 to 360. For each configuration, the framework determines the minimum required \emph{Operation Table} depth -- which directly relates to execution latency.
	
	\subsubsection{Comparison with Baselines} 
	We compare the proposed framework against three baseline partitioning strategies. (1) The post-neuron round-robin approach assigns fan-in synapses of each neuron to SPUs in round-robin order, ensuring each neuron's partial current is maintained within a single SPU and avoiding neuron state duplication. However, varying fan-in counts across neurons produce imbalanced synaptic workloads, increasing the required \emph{Operation Table} depth and execution latency. (2) The synapse round-robin strategy assigns individual synapses in round-robin fashion, achieving balanced workload distribution but requiring partial currents of neurons to be stored across all SPUs, causing widespread neuron state duplication that significantly increases \emph{Unified Memory} requirements. (3) The weight round-robin baseline clusters synapses sharing the same weight value and distributes these clusters across SPUs, maximizing weight reuse but tending to increase neuron duplication and workload imbalance.
	
\begin{figure}
		\centering
		\includegraphics[width=\linewidth]{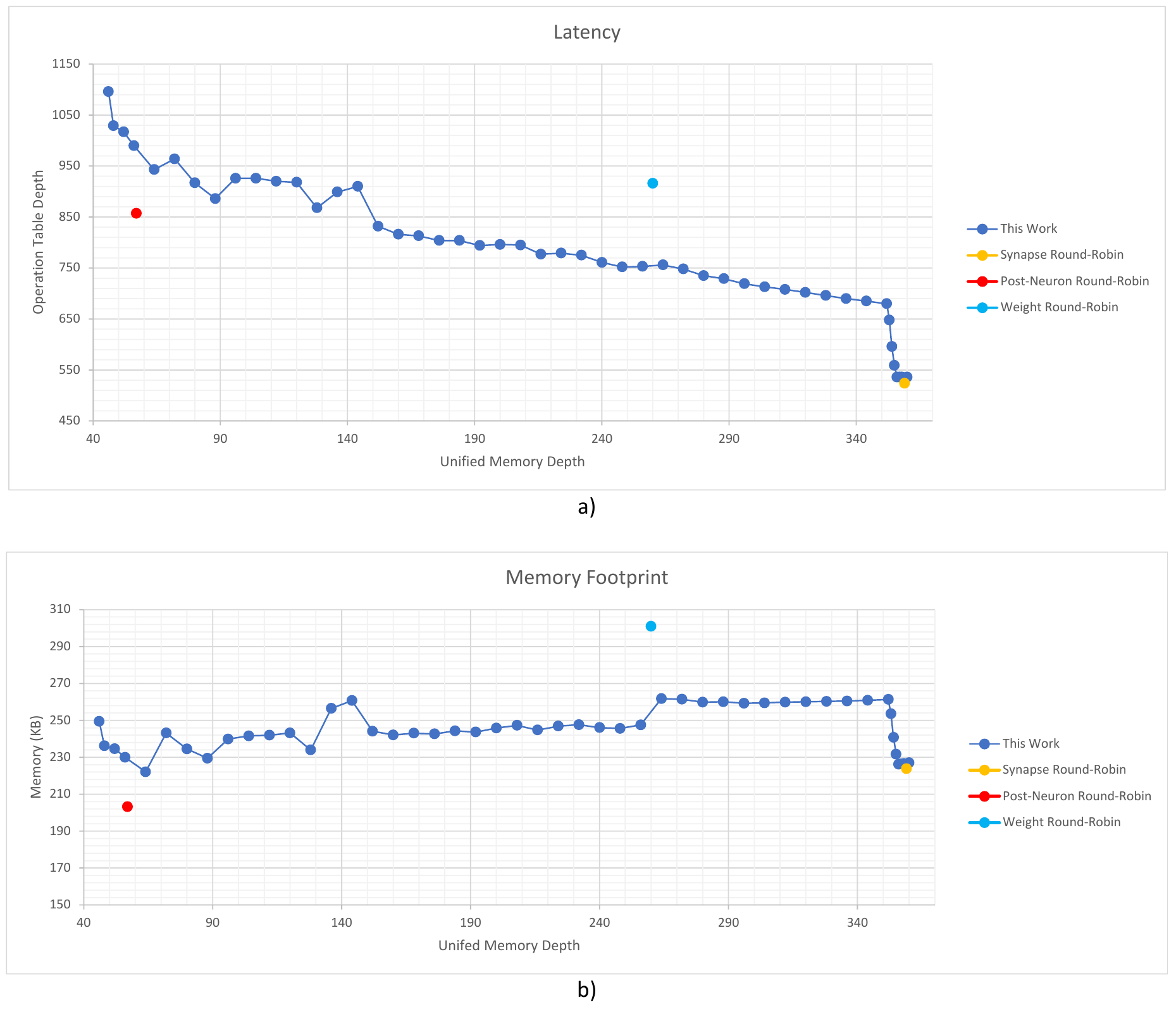}%
		\caption{(a) Minimum required \emph{Operation Table} depth determined by the proposed framework (directly related to inference latency) for different \emph{Unified Memory} depths, compared with three baseline solutions. (b) Required memory of the framework's  solutions for different \emph{Unified Memory} depths, compared with three baseline solutions.}
		\label{fig:partitioning_performance}
	\end{figure}

	Figure~\ref{fig:partitioning_performance}a illustrates the relationship between \emph{Unified Memory} depth and the minimum required \emph{Operation Table} depth for all approaches. The proposed framework begins from a fully balanced synaptic distribution and incrementally adjusts the mapping to satisfy the \emph{Unified Memory} constraint. As the constraint relaxes, the framework maintains a more balanced distribution, reducing the required \emph{Operation Table} depth and therefore latency. Specifically, at a \emph{Unified Memory} depth of 46, the framework requires an \emph{Operation Table} depth of 1096, while increasing the depth to 360 reduces this to 536.
	
	The synapse round-robin baseline requires a \emph{Unified Memory} depth of 359 and an \emph{Operation Table} depth of 539, nearly identical to the proposed framework under the same constraint, confirming that the framework converges to the best achievable balanced solution under relaxed memory conditions. The weight round-robin strategy requires a \emph{Unified Memory} depth of 260 with an \emph{Operation Table} depth of 916, approximately 17.5\% larger than the framework under a similar constraint, reflecting its less efficient workload distribution. The post-neuron round-robin approach performs strongly under tight constraints, requiring only a \emph{Unified Memory} depth of 57 with an \emph{Operation Table} depth of 857, compared with 990 by the proposed framework — a 15.5\% improvement. However, this advantage exists only under tight constraints: once the \emph{Unified Memory} depth exceeds approximately 144, all solutions produced by the proposed framework outperform post-neuron round-robin in latency, as the latter cannot adapt its workload balance to exploit additional available memory. Furthermore, the framework is capable of generating valid mappings for smaller \emph{Unified Memory} capacities (e.g., a depth of 46).
	
	Figure~\ref{fig:partitioning_performance}b compares total memory footprints. The proposed framework fluctuates between 222.12 KB and 261.32 KB across all configurations. The weight round-robin strategy produces a memory footprint of 300.91 KB -- approximately 15\% larger compared with the framework under similar constraint -- due to additional neuron duplication. The post-neuron round-robin approach achieves a footprint nearly 12\% lower than the framework at its optimal constraint point of depth 57, mirroring its latency advantage under the same condition. However, beyond this point, its \emph{Operation Table} memory capacity remains fixed despite increases in \emph{Unified Memory} depth, while the proposed framework continues reducing \emph{Operation Table} size by exploiting additional memory for better balance. Consequently, for \emph{Unified Memory} depths beyond 57, the post-neuron round-robin memory footprint becomes inferior to that of the framework.
	
	\subsubsection{Utilization Evaluation}
\begin{figure}
		\centering
		\includegraphics[width=\linewidth]{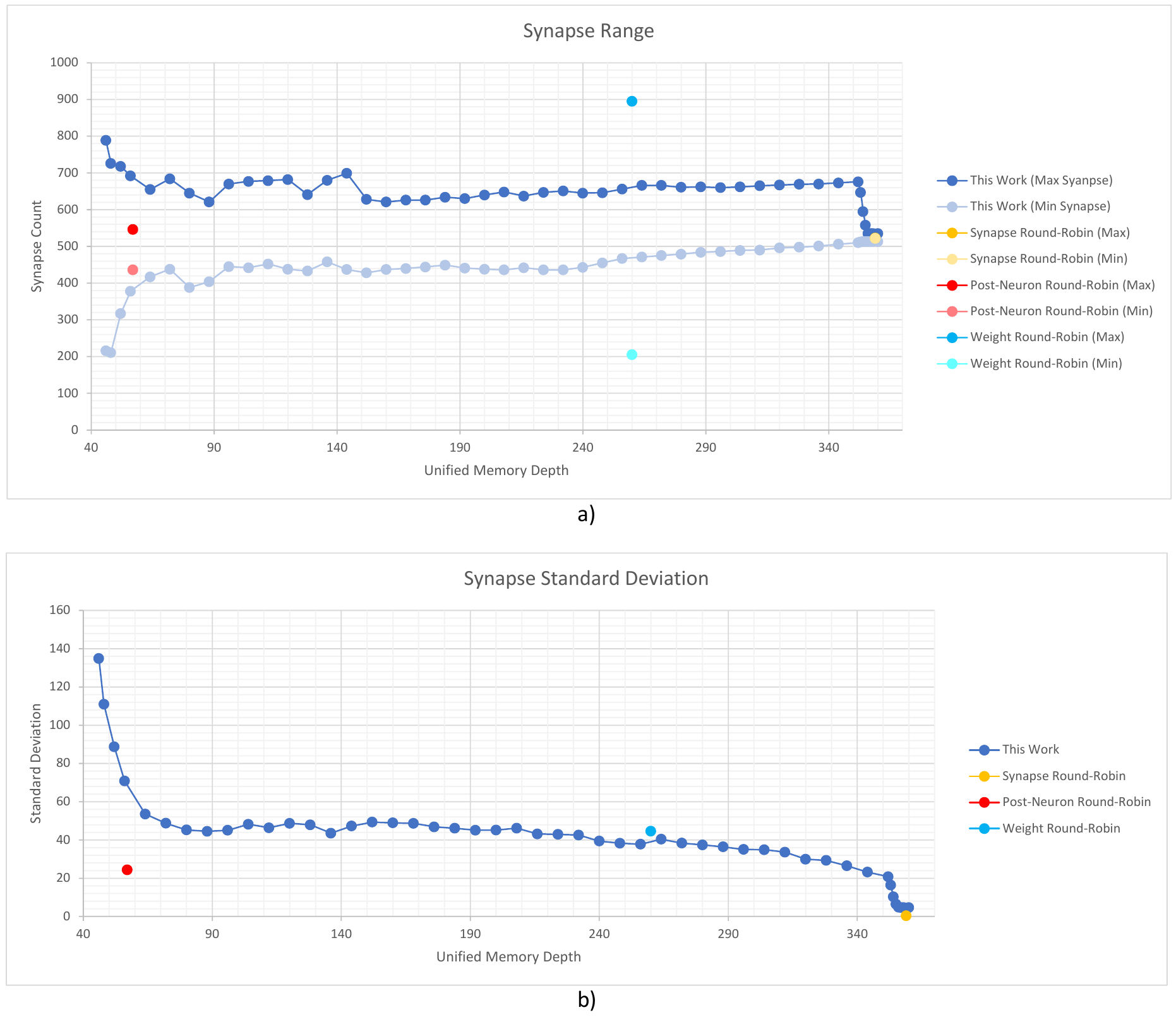}%
		\caption{(a) Maximum and minimum numbers of synapses mapped to a single SPU for different \emph{Unified Memory} depths. (b) Standard deviation of the number of synapses mapped to SPUs across different \emph{Unified Memory} depths.}
		\label{fig:Balanced_Distribution}
	\end{figure}

	\noindent In the next step, we evaluate how effectively the proposed framework balances the synaptic workload among SPUs. Figure~\ref{fig:Balanced_Distribution}a presents the maximum and minimum synapse counts per SPU across configurations. Under relaxed memory constraints, the synaptic workload remains nearly identical across SPUs, producing a minimal gap between maximum and minimum counts. As the constraint tightens, the framework introduces controlled imbalance to satisfy memory requirements, widening the gap between maximum and minimum synapse counts per SPU progressively. Figure~\ref{fig:Balanced_Distribution}b confirms this through the standard deviation of synapse counts, which steadily decreases as \emph{Unified Memory} depth increases, approaching zero under fully relaxed constraints and demonstrating convergence toward an almost perfectly balanced allocation.
	
	\subsubsection{Post-Neuron Centralization and Weight Reusability}
\begin{figure}
		\centering
		\includegraphics[width=\linewidth]{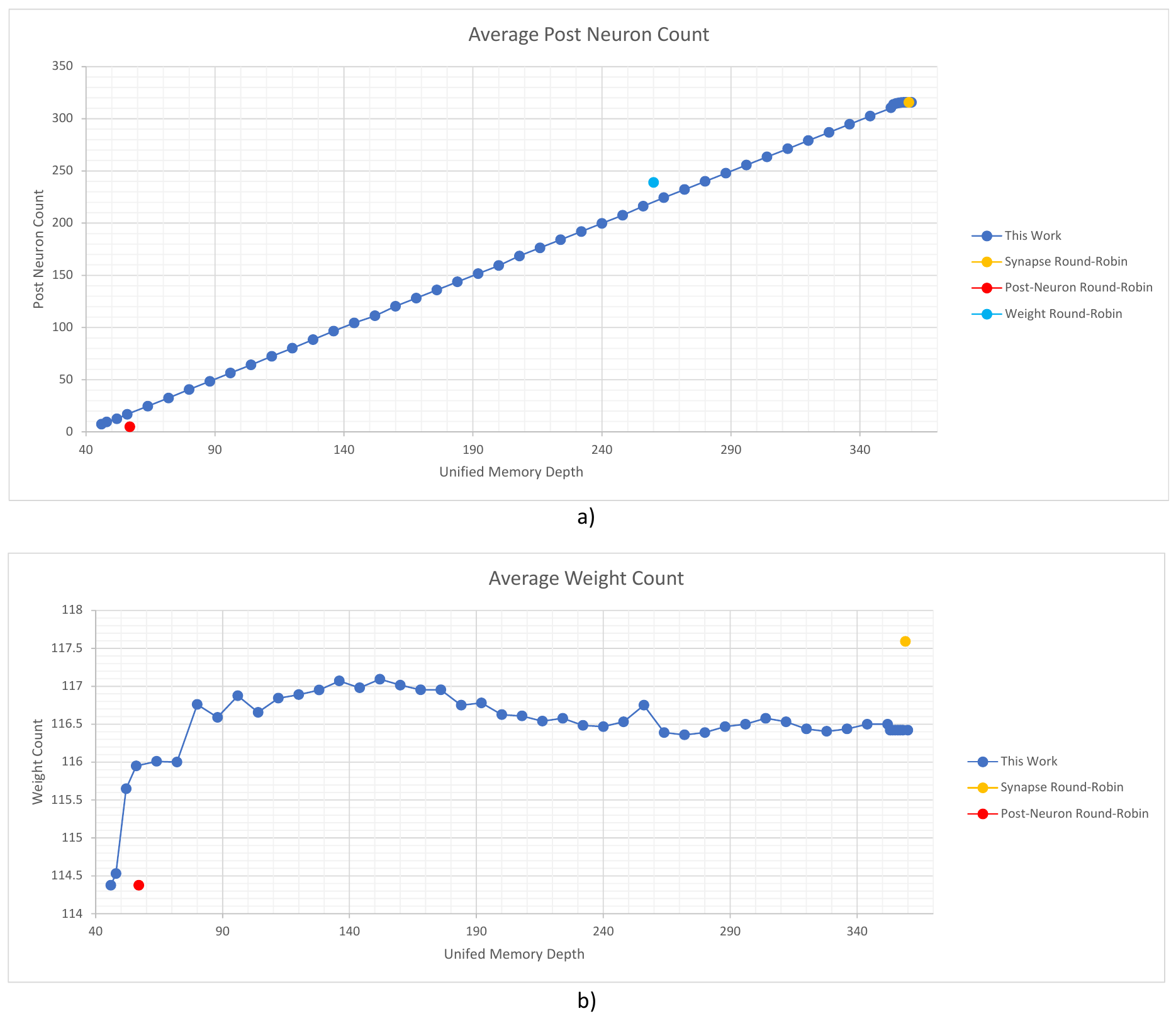}%
		\caption{(a) Average number of post-neurons mapped to each SPU for different \emph{Unified Memory} depths. (b) Average number of weights mapped to each SPU across different \emph{Unified Memory} depths.}
		\label{fig:memory_compression}
	\end{figure}

	\noindent Finally, we analyze how the proposed framework reduces post‑neuron duplication while simultaneously exploiting weight reusability to satisfy different \emph{Unified Memory} constraints. Figure~\ref{fig:memory_compression}a illustrates the average number of post-neurons assigned to each SPU across different \emph{Unified Memory} depths. This value grows steadily and nearly linearly as memory capacity increases, reflecting the framework's ability to distribute synapses more evenly when \emph{Unified Memory} depth permit, while limiting post-neuron duplication under tighter constraints to maintain feasible mappings.
	
	Figure~\ref{fig:memory_compression}b shows the average number of weights assigned to each SPU for the same set of configurations. For \emph{Unified Memory} depths larger than approximately 80, the average weight count remains relatively stable around 116.5, indicating that the framework maintains a consistent level of weight reuse under moderate and relaxed constraints. However, when the \emph{Unified Memory} depth becomes smaller than this threshold, the average weight count decreases noticeably to approximately 114.375, as the framework shifts priority toward weight reusability once post-neuron centralization reaches its limit. This behavior reflects the embedded priority rules within the partitioning phase: since storing a post-neuron consumes memory equivalent to $K$ weights, post-neuron sharing is prioritized first, with weight reuse becoming the key mechanism under the tightest constraints -- enabling the framework to produce valid mappings even below the minimum \emph{Unified Memory} depth required by the post-neuron round-robin strategy.
	
	In conclusion, these results demonstrate that the proposed framework systematically leverages post-neuron centralization and weight reusability to address the performance--memory trade-off, consistently outperforming all baselines under moderate and relaxed constraints while remaining viable under tight memory conditions.
	
	\section{Conclusion}
	\noindent This paper presents SupraSNN, a highly parallel and energy-efficient hardware accelerator, together with a software co-design framework for partitioning and scheduling. Inspired by superscalar processors, SupraSNN provides \emph{synapse-level parallelism} by decoupling synaptic and neuronal computation and exploiting network sparsity to improve both memory efficiency and computational performance through \emph{Operation-Style} execution in SPUs. The architecture integrates parallel SPUs for synaptic processing, a centralized Neuron Unit for state updates, a lightweight high-throughput MC tree for spike distribution, and a bufferless ME tree for merging partial results.
	To satisfy memory constraints while maintaining high throughput, the probabilistic partitioning software tool employs post‑neuron centralization and weight reusability to transform an initially performance‑optimal but memory‑intensive mapping into a memory‑efficient yet well-balanced solution. A complementary heuristic scheduling stage further enhances throughput by exploiting slack periods across post‑neuron workloads, effectively overlapping synaptic execution with partial‑result merging.
	This co-design strategy enables competitive performance on digit-classification benchmarks and achieves state-of-the-art results among fully connected SNN accelerators, including a latency of 0.149 ms/image and an energy consumption of 0.02563 mJ/image on MNIST, while also demonstrating reliable performance on SHD.
	Future work will focus on the primary bottleneck of SupraSNN--memory overhead--and incorporate temporal sparsity-aware mechanisms, a defining characteristic of SNNs, to enhance latency, scalability, and support for large-scale SNN deployments.

\end{document}